\documentclass{article}

\usepackage{amsmath}
\usepackage{amssymb}
\usepackage{graphicx}
\usepackage{hyperref}
\usepackage{longtable}
\usepackage{float}

\newcommand{\OP}[1]{\check{#1}}

\newcommand{\VC}[1]{\boldsymbol{#1}}

\newcommand{\UV}[1]{\hat{\boldsymbol{#1}}}

\newcommand{\BK}[2]{\langle #1 \mid #2 \rangle}

\newcommand{\BA}[1]{\langle #1 \mid}

\newcommand{\KT}[1]{\mid #1 \rangle}

\newcommand{\MX}[1]{\left[ #1 \right]}

\newcommand{\bibentry}[5]{#1, {#2} {\bf #3}, #4 (#5).}

\begin{document}

\title{$^{3}$H and $^{3}$He calculations without angular momentum decomposition}

\author{K. Topolnicki \\ Jagiellonian University, PL-30348 Krak{\'o}w, Poland}

\date{\today}

\maketitle

\begin{abstract}
Results for the three nucleon (3N) bound state
carried out using the ``three dimensional" (3D) formalism are
presented.
In this approach calculations are performed without the use
of angular momentum decomposition and instead rely directly on
the 3D degrees of freedom of the nucleons. In this paper, for the first time, 
3D results for $^{3}$He bound state with the inclusion of the screened
Coulomb potential are compared to $^{3}$H calculations. Additionally, using these
results, 
matrix elements of simple current operators related to the description of beta decay of the triton are
	given. 
All computations are
carried out using the first generation of NNLO two nucleon (2N) and 3N forces from the
Bochum - Bonn group.
\end{abstract}
\section{Introduction}

The ``three dimensional" approach is an alternative to
traditional, partial wave decomposition based, few-nucleon calculations. The main characteristic of the new
approach is that is does not rely on angular momentum decomposition and instead
works directly with the (\emph{three dimensional}) momentum degrees of freedom of the
nucleons. As a consequence, it is not necessary to numerically
implement the heavily oscillating functions needed for partial wave calculations. 
This beneficial property opens up the possibility to perform calculations at 
higher energies and with longer ranged potentials, including screened Coulomb
interactions. The latter is explored in this
paper for the $^{3}$He bound state. Some introductory applications of the 3D approach were
investigated by various research groups
\cite{o1,o2,a_new_treatment_of_2n_and_3n}. A very good,
general introduction to the 3D approach is given in \cite{three4}.

This paper is built on top of the work carried out in
\cite{a_three_dimensional,a_new_treatment_of_2n_and_3n,operator_form_of_3h}. 
Several important improvements were made with respect to the previous work presented
in \cite{a_three_dimensional}. 
An additional $\frac{3}{2}$ total isospin component is added 
and the 3D bound state calculations
are extended to the $^{3}$He nucleus by including a screened Coulomb interaction
from \cite{screened_coulomb}. The results presented in this paper
demonstrate that including 
potentials with a longer range (screened Coulomb interaction) is possible within
the 3D approach. This is
difficult in traditional calculations that utilize partial wave decomposition.
Another significant improvement is in the efficient implementation of the 3N
force in the calculations. This was previously, in \cite{a_three_dimensional}, 
a very numerically expensive
operation that required the numerical calculation of many-fold integrals and 
these integrals had to be calculated each time a new 3N energy was tested for the
existence of a bound state. In this paper a method is presented that allows the decoupling of
some of the integrals. In practice this means that a lot of the numerical work
can be performed once and the results of this work can be reused many times when
testing different energies for the existence of a bound state. After carrying out
the initial numerical work and storing the results, calculations with a 
3N force are not drastically more expensive then calculations with 2N
forces only.

The paper is organized as follows. In Sec. \ref{formalism} an introduction to the
3D bound state calculations is provided. The Faddeev equation
is introduced together with the operator form of the 3N bound and Faddeev
states. This operator form allows states to be defined via sets of scalar
functions of Jacobi momenta. These scalar functions are the main object of the
3D calculations and in Sec. \ref{formalism} 
the Faddeev equation is transformed into a linear equation in a space
spanned by the scalar functions. Section \ref{numerical_realization} contains
more information on the practical numerical realization of the calculations. 
Results of the calculation are presented and discussed in 
Sec. \ref{numerical_realization}. Finally Sec. \ref{summary} contains the
summary and outlook. 
Appendix \ref{spin_operators} and
\ref{explicit_form} show various details of the numerical implementation
(see also \cite{a_three_dimensional}) and Appendix \ref{trbetadecay} contains details
related to the calculation of current operator matrix elements related to the
trition beta decay.

\section{Formalism}
\label{formalism}

The starting point of the calculations is an operator form
of the 3N (Faddeev, bound) state from \cite{a_three_dimensional,operator_form_of_3h}. In this form a 3N state $\KT{\phi}$ can be written as:
\begin{align} \nonumber
	\BK{\VC{p} \VC{q}; (t \frac{1}{2}) T M_{T}}{\phi} = \sum_{i = 1}^{8}
	\phi^{(i)}_{tT}(p , q , \UV{p} \cdot \UV{q}) \\
	\OP{O}_{i}(\VC{p} , \VC{q})
	\KT{\chi^{m}}
	\label{state_operator_form}
\end{align}
where $\VC{p}$, $\VC{q}$ are Jacobi momenta; $t$ is the isospin in the two nucleon
subsystem; $T$ is the total isospin; $M_{T}$ is its projection
($-\frac{1}{2}$ for $^{3}$H, $\frac{1}{2}$ for $^{3}$He); the spin operators $\OP{O}_{i = 1 \ldots 8}(\VC{p} , \VC{q})$ are
listed in Appendix \ref{spin_operators} and $\KT{\chi^{m = \pm \frac{1}{2}}} = \KT{(0 \frac{1}{2})
\frac{1}{2} m}$ is a given spin state where the spins of the three nucleons are
coupled to the total spin $\frac{1}{2}$ with projection $m$.  
The set
of scalar functions $\phi^{(i)}_{tT}(p , q , \UV{p} \cdot \UV{q})$
in (\ref{state_operator_form}) for all indexes $i, t, T$
fully determines the
state $\KT{\phi}$. In
further parts of the paper a notation will be used in which this whole 
set of scalar functions 
is referred to by the Greek letter without the
indexes and arguments. 
For example:
the 3N states $\KT{\phi}, \KT{\gamma}, \KT{\beta}$, $\ldots$
are defined by sets of scalar functions (from the operator form
(\ref{state_operator_form})) $\{\phi^{(i)}_{tT}(p , q , \UV{p} \cdot
\UV{q})\}$, $\{\gamma^{(i)}_{tT}(p , q , \UV{p} \cdot \UV{q})\}$, 
$\{\beta^{(i)}_{tT}(p , q , \UV{p} \cdot \UV{q})\}$, $\ldots$ and these sets will be simply
refered to as $\phi, \gamma, \beta, \ldots$. Moreover, the scalar functions in $\phi,
\gamma, \beta, \ldots$ span a linear space that will form the main stage of the 3D
calculations and vectors from this space will also be referred to using the same greek 
letters $\phi, \gamma, \beta, \ldots$

Calculations of the 3N bound state are carried out within the Faddeev formalism. 
Results presented in this paper are obtained with a version of the Faddeev equation that was
investigated in \cite{a_three_dimensional} and that does not use
the 2N transition operator:
\begin{equation}
	\KT{\psi} = \OP{G}_{0}(E) \left( \OP{V} + \OP{V}^{(1)}
	\right) \left(\OP{1} + \OP{P} \right) \KT{\psi}.
	\label{faddeq}
\end{equation}
In (\ref{faddeq}) $\KT{\psi}$ is a Faddeev component, $\OP{G}_{0}(E)$ is
the free propagator for energy $E$, $\OP{V}$ is the two nucleon potential between
particles two and three and $\OP{V}^{(1)}$ is a part of the 3N force
that is symmetric with respect to the exchange of particles two and three.
Finally:
\begin{equation}
	\OP{P} = \OP{P}_{12} \OP{P}_{23} + \OP{P}_{13} \OP{P}_{23}
	\label{perop}
\end{equation}
is composed from operators $\OP{P}_{ij}$ that perform the exchange of two
particles $i = 1
\ldots 3 , j = 1 \ldots 3$ and
the full 3N bound state wave function $\KT{\Psi}$ can be obtained from the
Faddeev component $\KT{\psi}$ using:
\begin{equation}
	\KT{\Psi} = \left( \OP{1} + \OP{P} \right) \KT{\psi}.
\end{equation}

Plugging the operator form (\ref{state_operator_form}) into the Faddeev equation
(\ref{faddeq}) and removing the spin dependencies results in a new equation 
for the scalar functions that define the 3N state. This work was performed in
\cite{a_three_dimensional} and results in the following equation:
\begin{equation}
	\OP{A}(E) \psi = \psi
	\label{eq1}
\end{equation}
where $\OP{A}(E)$ is an energy dependent linear operator that acts in a space
spanned by the scalar functions from (\ref{state_operator_form}). In practical calculations a slightly modified
version of this equation:
\begin{equation}
	\OP{A}(E) \psi = \lambda \psi
	\label{eq2}
\end{equation}
is solved for various values of the energy $E$. Once a solution to (\ref{eq2}) is found such that
$\lambda = 1$ then $\psi$ is also a solution to (\ref{eq1}) and $E$ is the bound
state energy.

The operator $\OP{A}$ can be split into several parts that correspond to the
various operators that appear in (\ref{faddeq}). These parts correspond
to the application of $\OP{1} + \OP{P}$ ($\OP{A}_{1 + P}$), $\OP{G}_{0}(E)
\OP{V}$ ($\OP{A}_{G_{0} V}$), $\OP{G}_{0}(E)
\OP{V^{(1)}}$ ($\OP{A}_{G_{0} V^{(1)}}$) onto the 3N state (represented by
the set of scalar functions $\psi$). In $^{3}$He calculations, the two nucleon force $\OP{V}$ can be
further split into the short nucleon-nucleon interaction 
$\OP{V}_{\text{NN}}$ and the longer ranged screened Coulomb potential
$\OP{V}_{\text{C}}(R)$:
\begin{equation}
	\OP{G}_{0}(E) \OP{V} = \OP{G}_{0}(E) \OP{V}_{\text{NN}} + \OP{G}_{0}(E)
	\OP{V}_{\text{C}}(R)
\end{equation}
where $R$ is the screening radius. This results in the operator
$\OP{A}_{G_{0} V}$ 
being written as a sum of two parts that correspond to the short range nuclear
and long range Coulomb interaction:
\begin{equation*}
	\OP{A}_{G_{0} V}(E) = \OP{A}_{G_{0 }V_{NN}}(E) + \OP{A}_{G_{0} V_{C}}(E , R).
\end{equation*}
For $^{3}$He, the effect of this separation is four linear operators (acting in the linear space spanned by the scalar
functions) and an additional dependence on the screening radius $R$:
\begin{align} \nonumber
	\OP{A}(E) = \OP{A}(E , R) =  \\
	\left( \OP{A}_{\text{G}_{0}\text{V}_{NN}}(E) + 
	\OP{A}_{\text{G}_{0}\text{V}_{C}}(E , R) +
	\OP{A}_{\text{G}_{0}\text{V}^{(1)}}(E) \right) \OP{A}_{\text{1+P}}.
\end{align}
the $\OP{A}$ is slightly less complicated and there are only three
operators:
\begin{equation}
	\OP{A}(E) = \left( \OP{A}_{\text{G}_{0}\text{V}_{NN}}(E) + 
	\OP{A}_{\text{G}_{0}\text{V}^{(1)}}(E) \right) \OP{A}_{\text{1+P}}.
\end{equation}
The explicit form of these operators is given in Appendix \ref{explicit_form}. Apart
from a different notation and some different bookkeeping, these expressions
are very similar to those presented in
\cite{a_three_dimensional}. For this reason
their discussion is moved from the main text into Appendix \ref{explicit_form}.

In this paper we again solve Eq. (\ref{eq2}) but several significant changes with respect to \cite{a_three_dimensional} have been
added to the calculations. The first change is the addition of the screened
Coulomb interaction for $^{3}$He. Secondly, a third $t = 1, T =
\frac{3}{2}$ isospin state is included in all calculations on top of $t = 0, T =
\frac{1}{2}$ and $t = 1, T = \frac{1}{2}$, this change has the
most significant impact on $^{3}$He calculations that utilize proton - proton
interactions. Thirdly, a new numerical integration scheme was developed to deal with the
application of the 3N force in
$\OP{A}_{\text{G}_{0}\text{V}^{(1)}}(E)$. The new scheme allows 
applications of $\OP{A}_{\text{G}_{0}\text{V}^{(1)}}(E)$ on scalar functions
for a number of different energies
to be carried out with little numerical work. Details on this new
approach are given in Appendix \ref{3N_force}. Finally, all results in this paper are
obtained 
assuming charge dependence of the nucleon - nucleon interaction and
using different proton-proton and neutron-proton versions of the 2N force 
(the neutron-neutron interaction is taken as the strong proton-proton
potential).
This change resulted in a shift of the triton bound state energy. This shift was 
additionally verified, for a Hamiltonian without the 3N force, using traditional
partial wave calculations.

\section{Numerical realization and results}
\label{numerical_realization}

Finding the solution to (\ref{eq2}) requires the computation of a matrix
representation of $\OP{A}$ ($\OP{A}(E)$ for $^{3}$H or $\OP{A}(E , R)$ for
$^{3}$He). This is achieved using
Krylov subspace methods. The procedure starts with an
initial set of scalar functions $\phi_{0}$ that satisfy the symmetry conditions
($\phi^{(i)}_{0,tT}(p , q , \UV{p} \cdot \UV{q}) = \pm \phi^{(i)}_{0,tT}(p , q
, -\UV{p} \cdot \UV{q})$ where $0$ is the index of the initial set of scalar
functions $\phi_{0}$)
outlined in \cite{a_three_dimensional}. Next, using the Arnoldi algorithm
(see e.g. \cite{Arnoldi}), a finite sized linear space spanned by 
the following set of $N$ vectors:
\begin{equation}
	\{\phi_{0}, \phi_{1} = \OP{A} \phi_{0} , \phi_{2} = \OP{A} \OP{A} \phi_{0} , \ldots , \phi_{N-1} = \OP{A}^{N - 1} \phi_{0} \}.
	\label{arnoldi_basis}
\end{equation}
is constructed. Here the lower index does not correspond to $t,T$ from
(\ref{state_operator_form}) but only numbers the vectors in the basis. Simultaneously the matrix elements of $\OP{A}$ are calculated 
within this subspace:
\begin{equation}
(\phi_{i} , \OP{A} \phi_{j})
\label{scal_prod}
\end{equation}
thus producing a $N \times N$ matrix representation of $\OP{A}$:
\begin{equation}
\MX{\OP{A}}_{i = 0 \ldots N - 1 \, j = 0 \ldots N - 1} = (\phi_{i} , \OP{A} \phi_{j}).
\end{equation}
The Arnoldi algorithm requires the implementation of a scalar product  
between two sets of scalar functions (\ref{scal_prod}). A simple formula was used that involved
spherical integrals over the Jacobi momenta $\VC{p}$, $\VC{q}$ and a summation over the values of
$k$, $t$, and $T$ of the product of scalar function values
$\phi_{i , tT}^{(k)}(p , q , \UV{p} , \UV{q}) \times \phi_{j , tT}^{(k)}(p , q , \UV{p}
, \UV{q})$ ($i,j$ are vector numbers in (\ref{arnoldi_basis})).
In addition to the scalar product all that is required for the
Arnoldi algorithm is the 
implementation of a subroutine that performs the application of $\OP{A}$ on the
scalar functions. This routine performs the integrals 
related to the various parts of 
$\OP{A}$ numerically as described in Appendix \ref{permutation_operator}, \ref{2N_potential} and 
\ref{3N_force}.

In this paper
two values for $N = 80 , 110$ were used. This resulted in $80 \times 80$ and $110
\times 110$ dimensional matrix eigenequations that correspond to
(\ref{eq2}) 
and the Fortran LAPACK library was used to solve these equations. 
Once solved all eigenvalues were discarded except one, whose
value was
closest to $1$ (see $\lambda$ values given further in the text). The corresponding eignevector:
\begin{equation}
v = (v_{0} , \ldots , v_{N - 1})
\end{equation}
was used to reconstruct the
scalar functions from (\ref{eq2}) by simple summation:
\begin{equation}
\psi = \sum_{i = 0}^{N-1} v_{i} \phi_{i}.
\label{reconstruct}
\end{equation}
Solutions $\psi$ calculated for energies $E$ (in $A(E)$ for $^{3}$H and $A(E , R)$ in $^{3}$He calculations) 
such that $\lambda$ is sufficiently
close to $1$ are good approximations of the solutions to (\ref{eq1}).
This also means that they are good approximations to the scalar functions of the
Faddeev components of the 3N bound state 
and the bound state energy is $E$. 
Naturally, it is important to verify the obtained solutions and for this
reason plots of the scalar functions $\psi$ presented further in this paper also contain
the functions $\beta$ calculated by a single application of $\OP{A}$:
\begin{equation}
\beta = \OP{A} \psi.
\label{check}
\end{equation} 
A correct solution to (\ref{eq1}) implies $\psi = \beta$.

The scalar functions $\Psi$ for the full 3N bound state wave function are obtained from the
scalar functions for the Faddeev component $\psi$ with (for more details see
Appendix
\ref{permutation_operator}):
\begin{equation}
	\Psi = \OP{A}_{\text{1 + P}} \, \psi.
	\label{wave_scalarr}
\end{equation}
Using the identity:
\begin{equation}
	\left( \OP{1} + \OP{P} \right)^{2} \equiv 3 \left( \OP{1} +
\OP{P} \right)
	\label{check1}
\end{equation}an additional simple verification of the calculations can be
established.
Namely, when $\OP{A}_{\text{1 + P}}$ is applied to $\Psi$
producing new functions $\zeta$:
\begin{equation}
	\zeta = \OP{A}_{\text{1 + P}} \Psi
	\label{check2}
\end{equation}
then (\ref{check1}) implies that $\zeta = 3 \Psi$. To check this, all plots
of $\Psi$ show also $\frac{1}{3} \zeta$.

The scalar functions $\psi, \Psi , \beta , \zeta$ that were obtained as a result 
of the Arnoldi algorithm \cite{Arnoldi} require normalization. 
A method very similar to the one outlined in 
equation ($58$) from \cite{a_three_dimensional} was used. 
This formula can be easily
modified to require only the scalar functions for the full bound state
$\Psi$ and this approach was used for all scalar functions. 

In the numerical realization scalar functions from
(\ref{state_operator_form}) are represented using multidimensional arrays.
Since there are $3$ isospin states:
\begin{align*}
\KT{(t = 0 \frac{1}{2}) T = \frac{1}{2} M_{T}} , \\
\KT{(t = 1 \frac{1}{2}) T = \frac{1}{2} M_{T}} , \\
\KT{(t = 1 \frac{1}{2}) T = \frac{3}{2} M_{T}} , 
\end{align*}
and $i = 1 , \ldots , 8$, a set of scalar functions 
$\phi \equiv \{\phi^{(i)}_{tT}(p , q , \UV{p} \cdot \UV{q})\}$ can be
represented using a $3 \times 8 \times N_{p} \times N_{q} \times N_{\UV{p} \cdot
\UV{q}}$ dimensional array where $N_{p}$, $N_{q}$ are the numbers of grid
points for the Jacobi momentum magnitudes and $N_{\UV{p} \cdot \UV{q}}$ is the number
of grid points for the angle between the Jacobi momenta.  

Two different values for the
number of lattice points $N_{p}$, $N_{q}$, $N_{\UV{p} \cdot \UV{q}}$ were used in calculations presented in this paper in
what will be reffered to as the \emph{first} and \emph{second} run. In
the first run of the code $N_{p} = N_{q} = N_{\UV{p} \cdot \UV{q}} = 16$ was
used and in the second run $N_{\UV{p} \cdot \UV{q}}$ was changed to $17$. 
This change was dictated by the different computer architecture used in
the second set of calculations.

The numerical calculation of integrals appearing in $\OP{A}$ (see Appendix
\ref{permutation_operator}, \ref{2N_potential} and \ref{3N_force}) and scalar products required by the Arnoldi
algorithm \cite{Arnoldi} creates the necessity of utilizing powerful computing resources. 
The first set of calculations was performed on the JUQUEEN cluster at the
J{\"u}lich Supercomputing Center (JSC) \cite{juqueen}. After the decommissioning of JUQUEEN, a
second set of calculations was performed on the new JURECA Booster module 
\cite{jureca_booster}. The KNL
architecture of the nodes available on the this new machine required an odd number
of points for $N_{\UV{p} \cdot \UV{q}}$. 

Additionally, the efficient utilization of the KNL architecture 
requires two-stage parallelization - using both the MPI protocol 
and OpenMP. During the second run, only the same pure MPI code 
that was used during the first set of calculations was available. This put a limit on
the number of grid points $N_{p}, N_{q}, N_{\UV{p} \cdot \UV{q}}$. 
Not being able to significantly increase the number of grid points in the second run,
in order to raise the precision of the calculations, the number of 
Gaussian integration points
for the azimuthal angle in (\ref{2N_integral}) was increased from $16$ to $64$, the 
number of iterations in the Arnoldi algorithm was increased from $N = 80$ to $N=110$
and the range of lattice points for $q$ was changed from $q <  5 [\text{fm}^{-1}]$ to
$q < 4 [\text{fm}^{-1}]$. The range of lattice points for $p$ remained unchanged 
$p <  5 [\text{fm}^{-1}]$. These combined changes resulted in a visible 
increase in the quality
of the results as will be shown in plots presented in this section. 
The currently obtained results also encourage the adaptation of two-stage parallelization,
but this is reserved for future calculations that will be performed with
more modern chiral potentials.

The calculations presented here use the same, first generation, chiral NNLO 2N and 3N potentials 
as were used in \cite{a_three_dimensional}. In this paper, however, both the proton-proton and neutron-proton
versions of the 2N force are used. 

\begin{figure}[H]
	\begin{center}
		\includegraphics[width = 0.8 \textwidth]{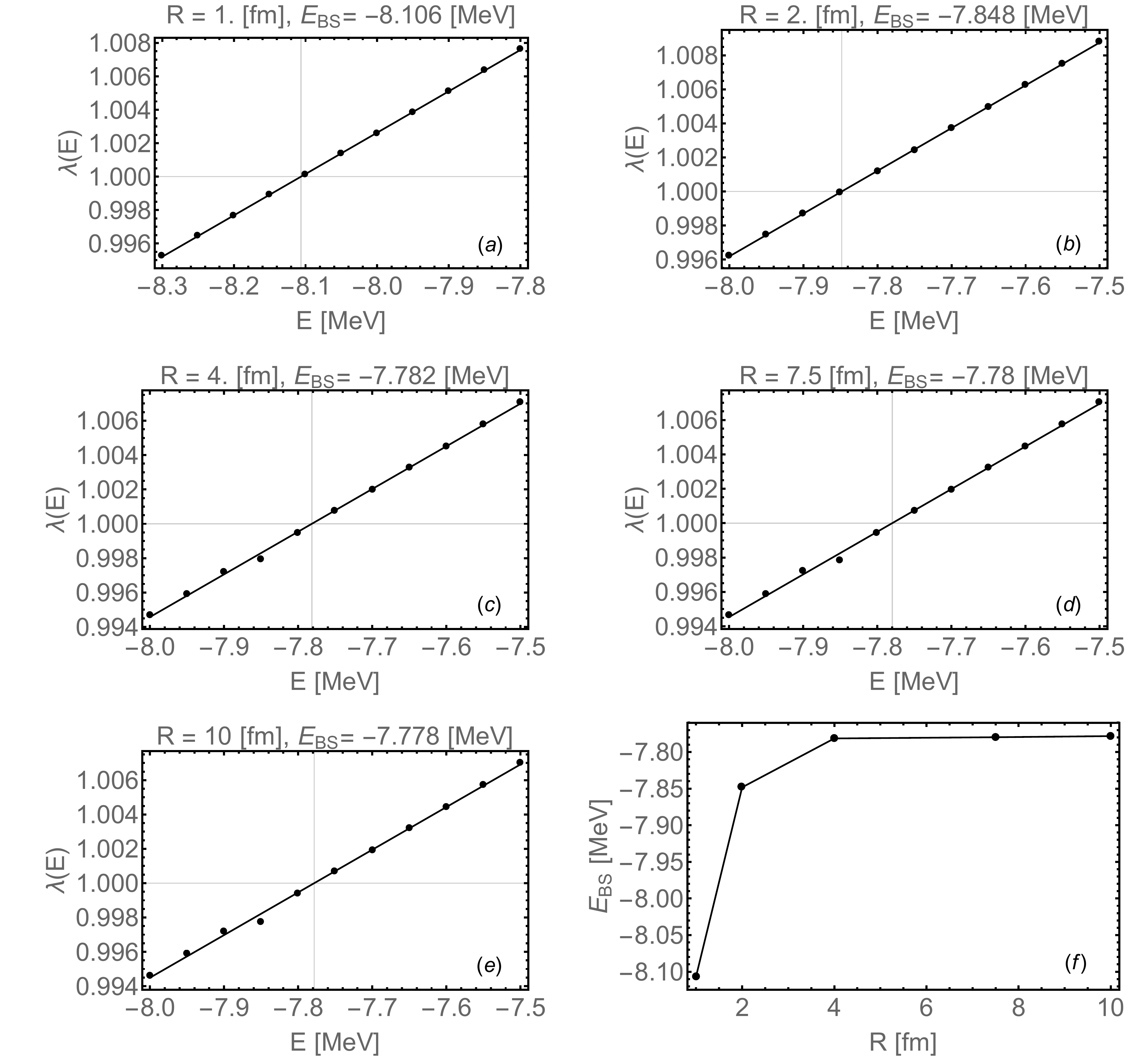}
	\end{center}
	\caption{
	$(a)$ - $(e)$: The eigenvalue $\lambda$ 
	that is closest to $1$ 
	as a functions of the energy $E$.
	The screened Coulomb potential from \cite{screened_coulomb} was used and
	consecutive plots correspond to different 
	screening radii used in equation (\ref{eq2}). 
	The crosses indicate the extrapolated bound state energies $E_{BS}$ and $\lambda = 1$.
	$(f)$: The extrapolated bound state energy $E_{BS}$ as a function of the
	screening radius $R$.
	Calculations were perfomed during the first run (see text) and use both the 2N and 3N force. 
	}
	\label{enplt}
\end{figure}

Figure \ref{enplt} shows the dependence of the $^{3}$He bound state
energy on the screening radius of the screened Coulomb interaction from
\cite{screened_coulomb}. Plots $(a)$ - $(e)$ show the eigenvalue $\lambda$
from (\ref{eq2}) that is closest to $1$ as a function of the energy $E$. These five
plots correspond to screening radii $R = 1 , 2 , 4 , 7.5 , 10 [\text{fm}]$ and
the data was obtained during the first run. The $\lambda(E)$ dependence is
visibly linear and a simple linear fit was used to extrapolate the bound state
energy $E_{BS}$ such that $\lambda(E_{BS}) = 1$. The $E_{BS}$ dependence on the
screening radius $R$ is
shown on plot $(f)$. The difference between
the extrapolated bound state energy for $R = 7.5 [\text{fm}]$ and $R = 10
[\text{fm}]$ is minimal and in the second run of the calculations a single value $R =
10 [\text{fm}]$ was used. The screened Coulomb potential from  
\cite{screened_coulomb} is a piece wise function, whose domain is divided into
three parts $0 - R$, $R - 2 R$, $2R - 3R$ and it reaches 
zero at distances greater then $3 R = 30 [\text{fm}]$. In the second run a
similar procedure was used to obtain the scalar functions for the bound states.
A range of energies was scanned looking for $\lambda = 1$, next a linear fit was
used to extrapolate to the bound state energy.

\begin{figure}[H]
	\begin{center}
		\includegraphics[width = 0.8 \textwidth]{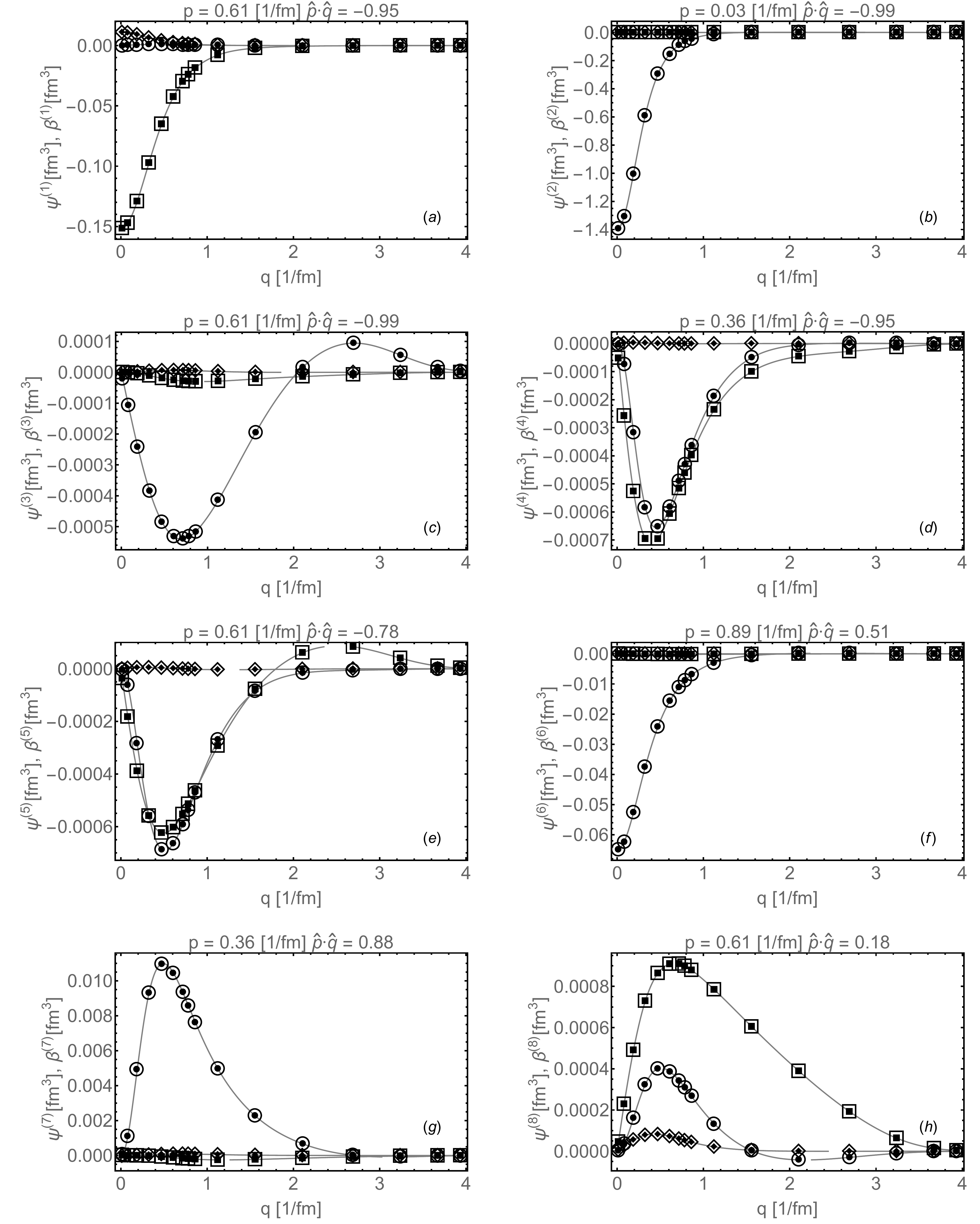}
	\end{center}
	\caption{Selected scalar functions for the Faddeev component of the $^{3}$He
	bound state plotted as a function of the Jacobi momentum $\VC{q}$ magnitude. 
	Scalar functions obtained directly using (\ref{reconstruct}) are marked using 
	solid markers while functions obtained using (\ref{check}) are marked using empty 
	markers. Squares circles and diamonds correspond to the three
	isospin states: $t=0,T=\frac{1}{2}$; 
	$t=1,T=\frac{1}{2}$ and $t=1,T=\frac{3}{2}$ respectively.  
	Calculations were performed during the second run (see text) using both 2N
	and 3N nuclear forces and a screened Coulomb interaction form
	\cite{screened_coulomb} with screening radius $R = 10 [\text{fm}]$ (the
	potential goes to zero at distances greater then $3 R = 30 [\text{fm}]$).
	The eigenvalue from (\ref{eq2}) for energy $E = -7.72654 [\text{MeV}]$ was $\lambda = 0.99997$.
	}
	\label{selected_scalar}
\end{figure}

\begin{figure}[H]
	\begin{center}
		\includegraphics[width = 0.8 \textwidth]{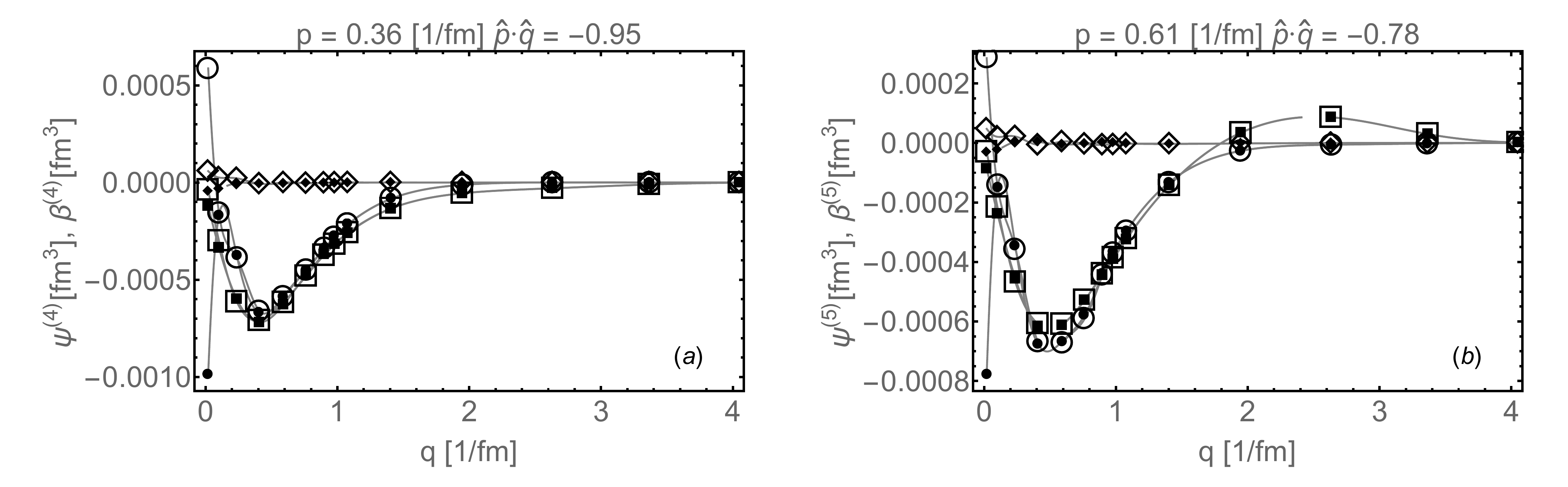}
	\end{center}
	\caption{Similar to $(d)$, $(e)$ from Fig. \ref{selected_scalar} but the scalar
	functions were obtained during the first run (see text). 	
	Calculations were carried out using both 2N
	and 3N nuclear forces and a screened Coulomb interaction form
	\cite{screened_coulomb} with screening radius $R = 10 [\text{fm}]$ (the
	potential goes to zero at distances greater then $3 R = 30 [\text{fm}]$).
	The eigenvalue from (\ref{eq2}) for energy $E = -7.77833 [\text{MeV}]$ was $\lambda = 0.99996$.
	}
	\label{selected_compare}
\end{figure}

Figure \ref{selected_scalar} shows selected scalar functions for the
Faddeev
component of the $^{3}$He bound state obtained during the second run. 
All plots contain both the $\psi$
functions from (\ref{eq1}) (that were reconstructed using (\ref{reconstruct})) 
and the $\beta$ functions from (\ref{check}). The
obtained values of these functions practically overlap, verifying the
solution. The additional $T = \frac{3}{2}$ component is visible only in the
dominant, first, scalar function on plot $(a)$. In Fig. \ref{selected_compare} 
selected scalar functions obtained during the first run are shown. When these two
plots, $(a)$ and $(b)$, are compared with plots $(d)$ and $(e)$ from Fig. \ref{selected_scalar}
numerical artifacts are visible for low values of $q$. The disappearance of
these artifacts in the second run is indicative of the positive effects of the
changes made in the second run of calculations. It should, however, be noted that the
numerical artifacts observed in the first run affect only the non-dominant
scalar functions whose values are relatively very small. For this reason they
are not expected to have a significant impact on the extrapolated bound state
energy and other observables.

\begin{figure}[H]
	\begin{center}
		\includegraphics[width = 0.8 \textwidth]{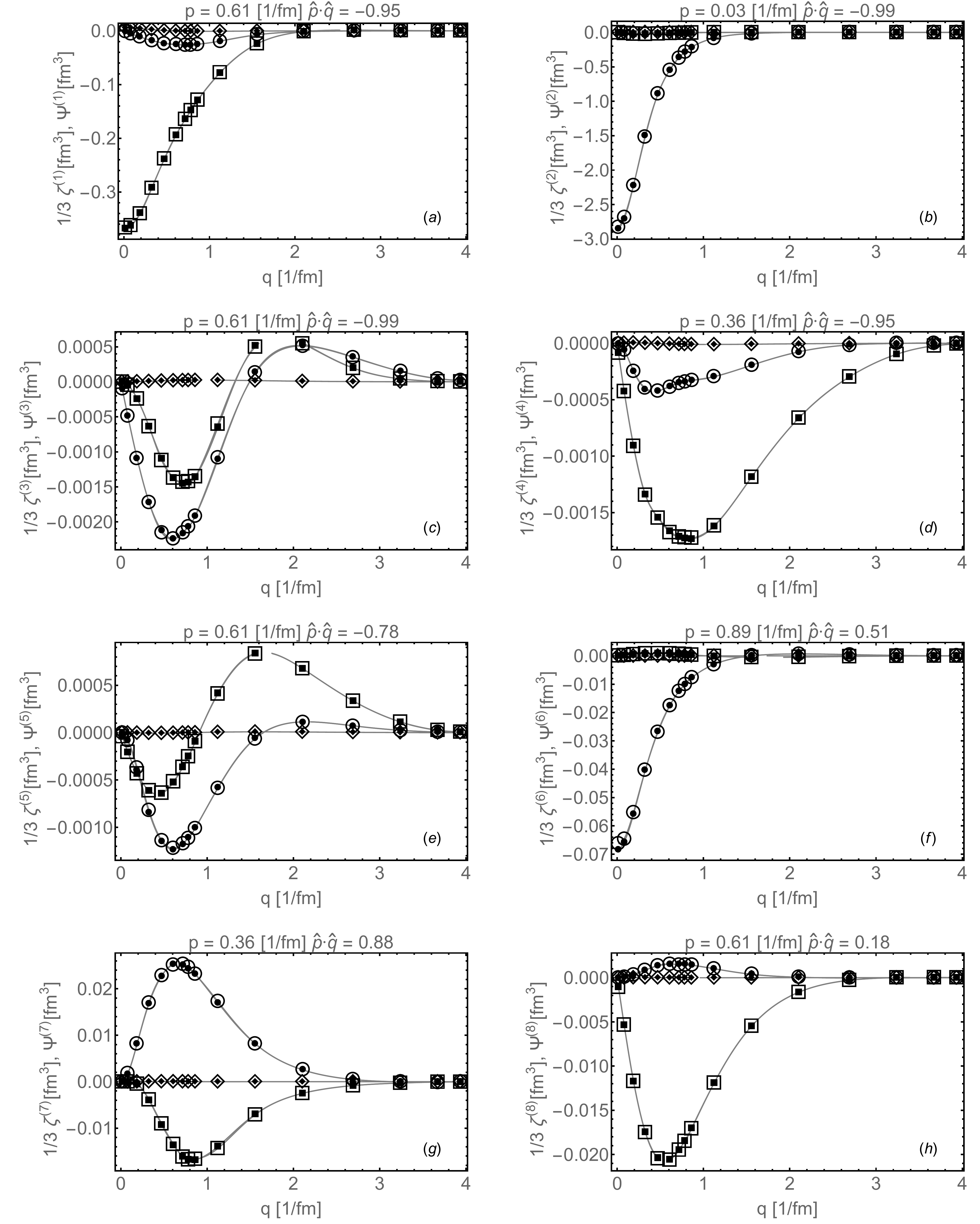}
	\end{center}
	\caption{Selected scalar functions for the $^{3}$He
	bound state. 
	Scalar functions obtained using (\ref{wave_scalarr}) are marked using solid 
	markers while functions obtained using (\ref{check2}) are marked using 
	empty markers. Squares, circles and diamonds respectively denote the three
	isospin states: $t=0,T=\frac{1}{2}$; 
	$t=1,T=\frac{1}{2}$ and $t=1,T=\frac{3}{2}$.  
	Calculations were performed during the second run (see text) using both 2N
	and 3N nuclear forces and a screened Coulomb interaction form
	\cite{screened_coulomb} with screening radius $R = 10 [\text{fm}]$ (the
	potential goes to zero at distances greater then $3 R = 30 [\text{fm}]$).
	The eigenvalue from (\ref{eq2}) for energy $E = -7.72654 [\text{MeV}]$ was $\lambda = 0.99997$.
	}
	\label{wave_scalar}
\end{figure}

\begin{figure}[H]
	\begin{center}
		\includegraphics[width = 0.6 \textwidth]{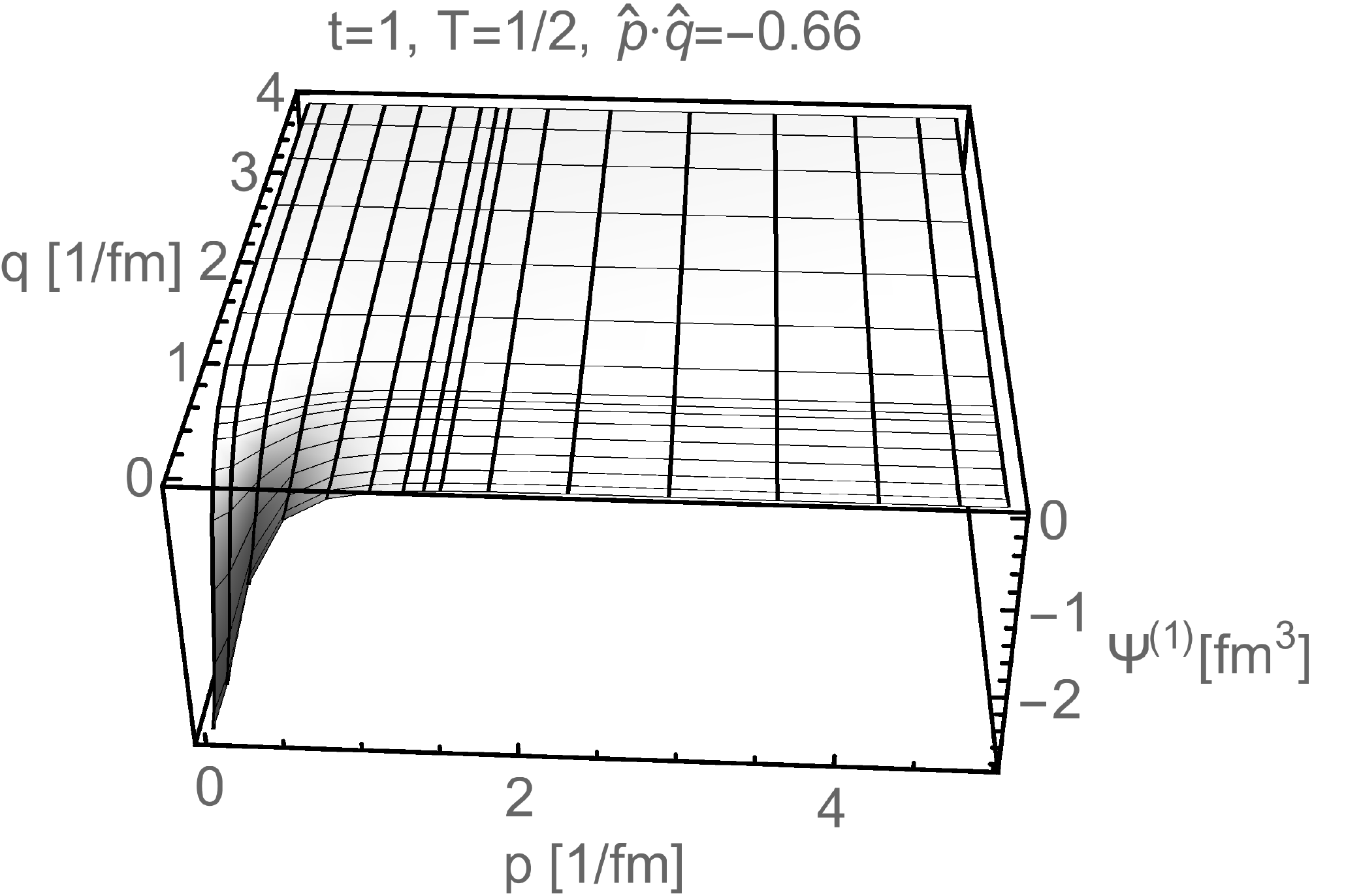}
	\end{center}
	\caption{Dominant scalar function for the $^{3}$He bound state plotted as a function
	of the magnitude of the Jacobi momenta $p$, $q$ for a chosen value of $\UV{p}
	\cdot \UV{q}$. 
	}
	\label{threedim3He}
\end{figure}

Figure \ref{wave_scalar} shows selected scalar functions $\Psi$ for the full $^{3}$He
bound state obtained using (\ref{wave_scalarr}) during the second run. All plots
also contain the $\frac{1}{3} \zeta$ functions from (\ref{check2}). The values
of these two functions essentially overlap again verifying the numerical
realization. 
Differences appear for non-dominant scalar functions and at low values of the 
momenta and angles. These differences can be attributed to problems with
interpolations in regions close to the function domain boundaries and 
are not visible 
in Fig.
\ref{wave_scalar} where the same values of $p$ and $\UV{p} \cdot \UV{q}$ were used as
in Fig. \ref{selected_scalar} for the purpouse of comparison. 
Figure \ref{threedim3He} exemplifies the dependence of
the dominant, first, scalar function $\Psi^{(1)}$ for the full $^{3}$He
bound state on the magnitude of both Jacobi momenta $p, q$ for a given angle
$\UV{p} \cdot \UV{q}$. It can be observed
that the scalar function quickly drops to zero in a region where the momenta are
greater then $\approx 2 [\frac{1}{\text{fm}}]$.

\begin{figure}[H]
	\begin{center}
		\includegraphics[width = 0.8 \textwidth]{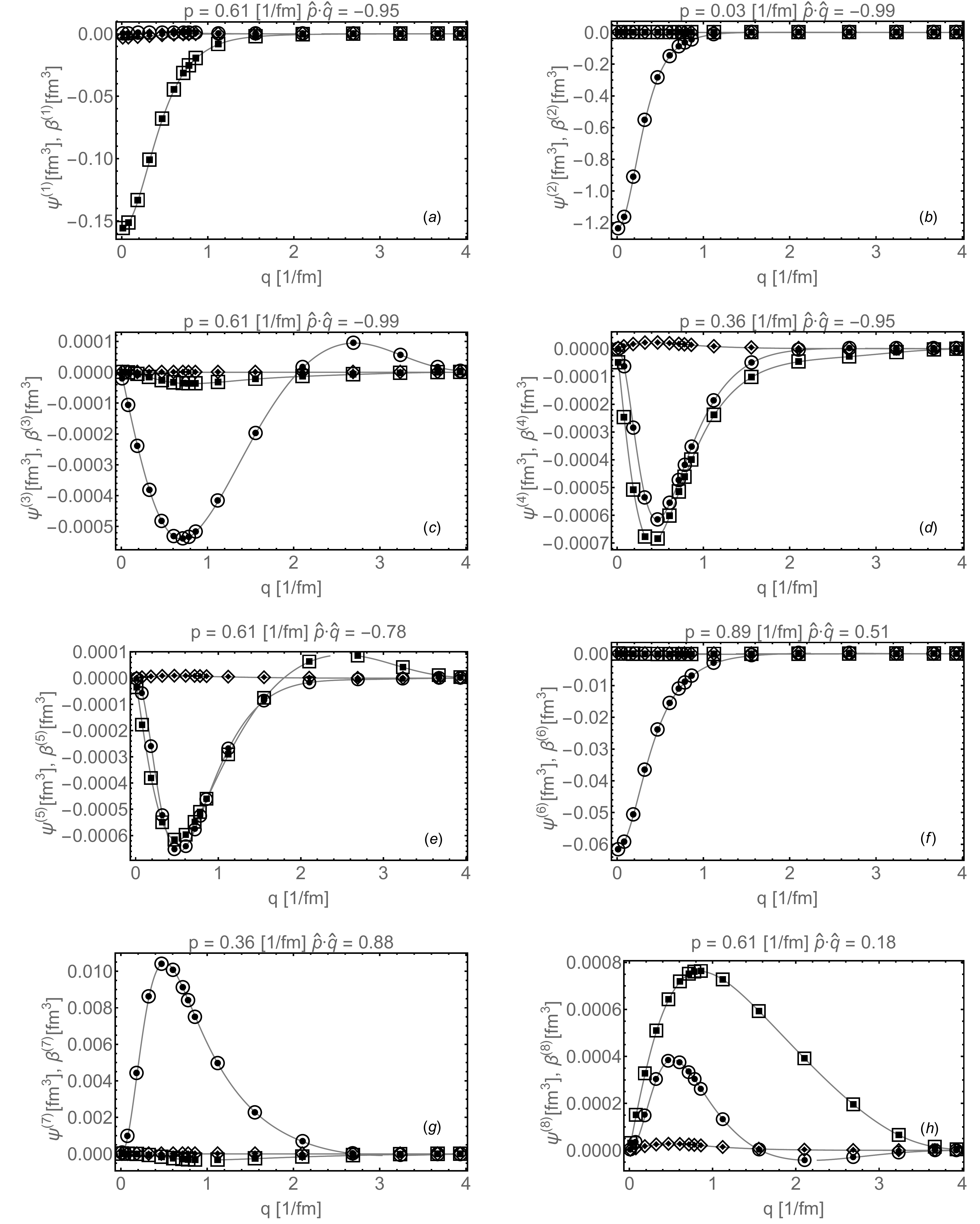}
	\end{center}
	\caption{
	Same as in Fig. \ref{selected_scalar} but for $^{3}$H.
	The eigenvalue from (\ref{eq2}) for energy $E = -8.4043 [\text{MeV}]$ was $\lambda = 0.99998$.
	}
	\label{selected_scalar_3H}
\end{figure}

\begin{figure}[H]
	\begin{center}
		\includegraphics[width = 0.8 \textwidth]{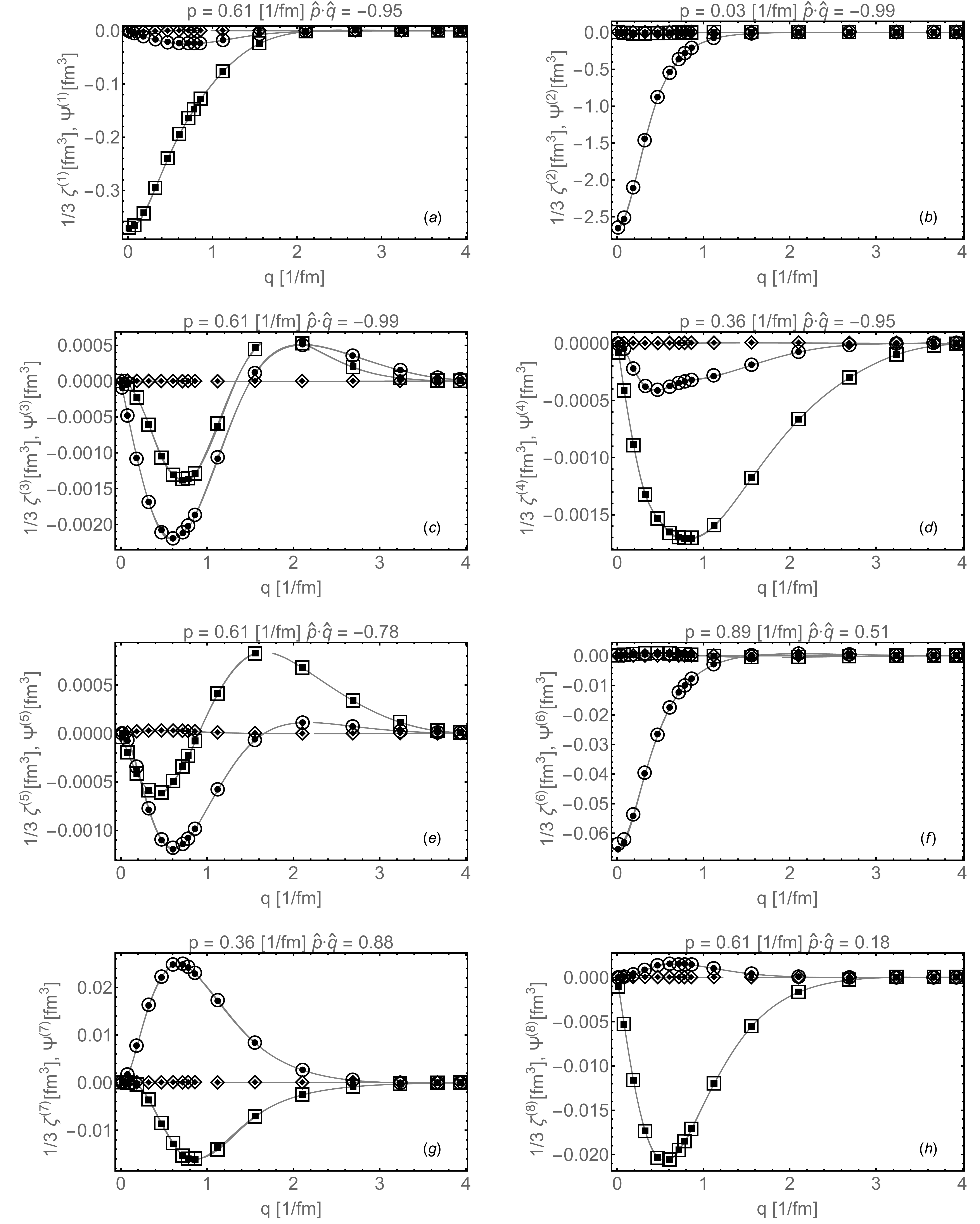}
	\end{center}
	\caption{Same as in Fig. \ref{wave_scalar} but for $^{3}$H.
	The eigenvalue from (\ref{eq2}) for energy $E = -8.4043 [\text{MeV}]$ was $\lambda = 0.99998$.
	}
	\label{wave_scalar_3H}
\end{figure}

\begin{figure}[H]
	\begin{center}
		\includegraphics[width = 0.8 \textwidth]{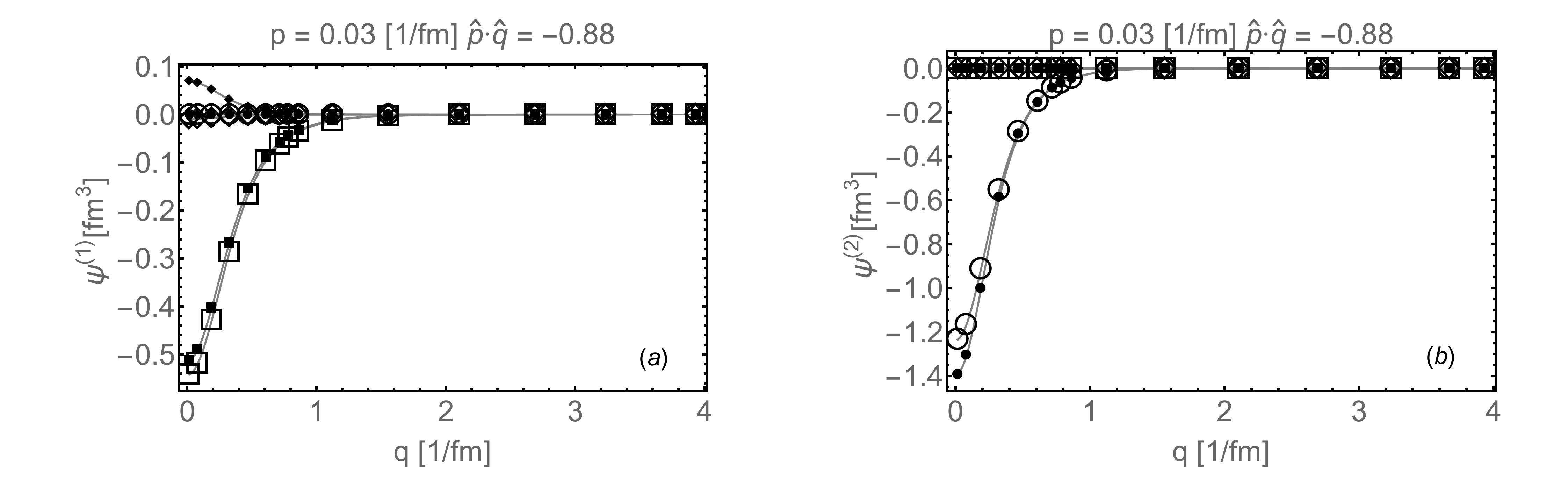}
	\end{center}
	\caption{Similar to Figure \ref{selected_scalar} and \ref{selected_scalar_3H}.
	Comparison of two dominant scalar functions ($(a)$ : $\psi^{(1)}$, $(b)$ : $\psi^{(2)}$) for the Faddeev component of
	$^{3}$H and $^{3}$He. Squares, circles, and diamonds denote the three
	isospin states: $t=0,T=\frac{1}{2}$; 
	$t=1,T=\frac{1}{2}$ and $t=1,T=\frac{3}{2}$ respectively. Empty markers correspond to
	$^{3}$H and solid markers correspond to $^{3}$He.
	}
	\label{wave_scalar_3H_c}
\end{figure}

\begin{figure}[H]
	\begin{center}
		\includegraphics[width = 0.8 \textwidth]{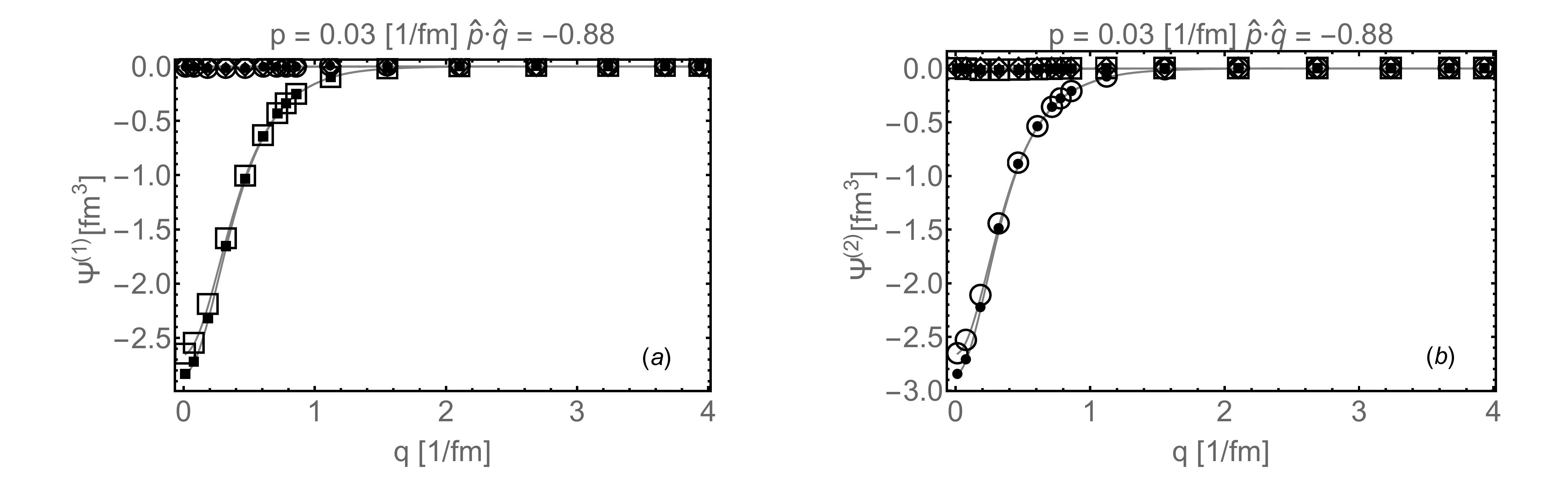}
	\end{center}
	\caption{Similar to Figure \ref{wave_scalar} and \ref{wave_scalar_3H}.
	Comparison of two dominant scalar functions ($(a)$ : $\Psi^{(1)}$, $(b)$ : $\Psi^{(2)}$) for the
	$^{3}$H and $^{3}$He bound state. Squares, circles, and diamonds denote the three
	isospin states: $t=0,T=\frac{1}{2}$; 
	$t=1,T=\frac{1}{2}$ and $t=1,T=\frac{3}{2}$ respectively. Empty markers correspond to
	$^{3}$H and solid markers correspond to $^{3}$He.
	}
	\label{wave_scalar_3H_c1}
\end{figure}

Analogous results for $^{3}$H are shown in Figs. \ref{selected_scalar_3H} and
\ref{wave_scalar_3H}. The differences between the two dominant Faddeev component
scalar functions $\psi$ are shown in Fig. \ref{wave_scalar_3H_c} for $^{3}$He
and $^{3}$H. The disappearance of
the $T = \frac{3}{2}$ component can be clearly seen when going from $^{3}$He to
$^{3}$H. In both $(a)$ and $(b)$ the $T = \frac{1}{2}$ components are dominant.
This dominance is also clearly visible in Fig. \ref{wave_scalar_3H_c1} where $(a)$
and $(b)$ contain the largest scalar functions $\Psi$ for the full wave function of $^{3}$H and
$^{3}$He.

Tables \ref{avg_1} and \ref{avg_2} contain expectation values of the
kinetic, 2N and 3N potential energies, for $^{3}$He and $^{3}$H, respectively.
The sum of these expectation values is close to the extrapolated bound state
energy. Differences can be attributed to the small number of grid points used to represent
the scalar functions. The values in these tables were obtained using the
expressions from \cite{a_three_dimensional}. The numerical integrals necessary
to calculate the expectation values of the 3N force are very similar
to the ones necessary to implement $\OP{A}_{G_{0} V^{(1)}}$ in \ref{3N_force}.
This made it possible to benefit from using the new parametrization of the
integrals that is described in Appendix \ref{3N_force} also when calculating the
expectation value of the 3N force.

\begin{table}[H]
\[
\begin{array}{ccccc}
 <E_{\text{kin}}> & <E_{\text{pot}}^{2N}> &
   <E_{\text{C}}^{2N}> & <E_{\text{pot}}^{3N}> &
   \Sigma  \\
32.36 & -39.91 & 0.6929 & -0.774 & -7.628 \\
\end{array}
\]
	\caption{Expectation values (in $[\text{MeV}]$) for the 
kinetic energy ($E_{\text{kin}}$), 
the strong 2N potential ($E_{\text{pot}}^{2N}$), 
the screened Coulomb interaction ($E_{\text{C}}^{2N}$), 
and the 3N force ($E_{\text{pot}}^{3N}$). These values 
were obtained using the methods outlined in \cite{a_three_dimensional}
from $^{3}$He bound state scalar functions obtained during the 
second run (see text).
The bound state was calculated using both 2N and 3N forces. 
The eigenvalue from (\ref{eq2}) for energy $E = -7.72654 [\text{MeV}]$ was $\lambda = 0.99997$.
} 
	\label{avg_1}
\end{table}

\begin{table}[H]
\[
\begin{array}{cccc}
 <E_{\text{kin}}> & <E_{\text{pot}}^{2N}> &
   <E_{\text{pot}}^{3N}> & \Sigma  \\
 32.91 & -40.45 & -0.7848 & -8.329 \\
\end{array}
\]
\caption{Expectation values (in $[\text{MeV}]$) for the 
kinetic energy ($E_{\text{kin}}$), 
the strong 2N potential ($E_{\text{pot}}^{2N}$), 
and the 3N force ($E_{\text{pot}}^{3N}$). These values 
were obtained using the methods outlined in \cite{a_three_dimensional}
from $^{3}$H bound state scalar functions obtained during the 
second run (see text). 
The bound state was calculated using both 2N and 3N forces. 
The eigenvalue from (\ref{eq2}) for energy $E = -8.4043 [\text{MeV}]$ was $\lambda = 0.99998$.
Additionally, the expectation value of the 2N screened Coulomb interaction was calculated:
$0.7054 [\text{MeV}]$.
} 
	\label{avg_2}
\end{table}

\begin{figure}[H]
	\begin{center}
		\includegraphics[width = 0.8 \textwidth]{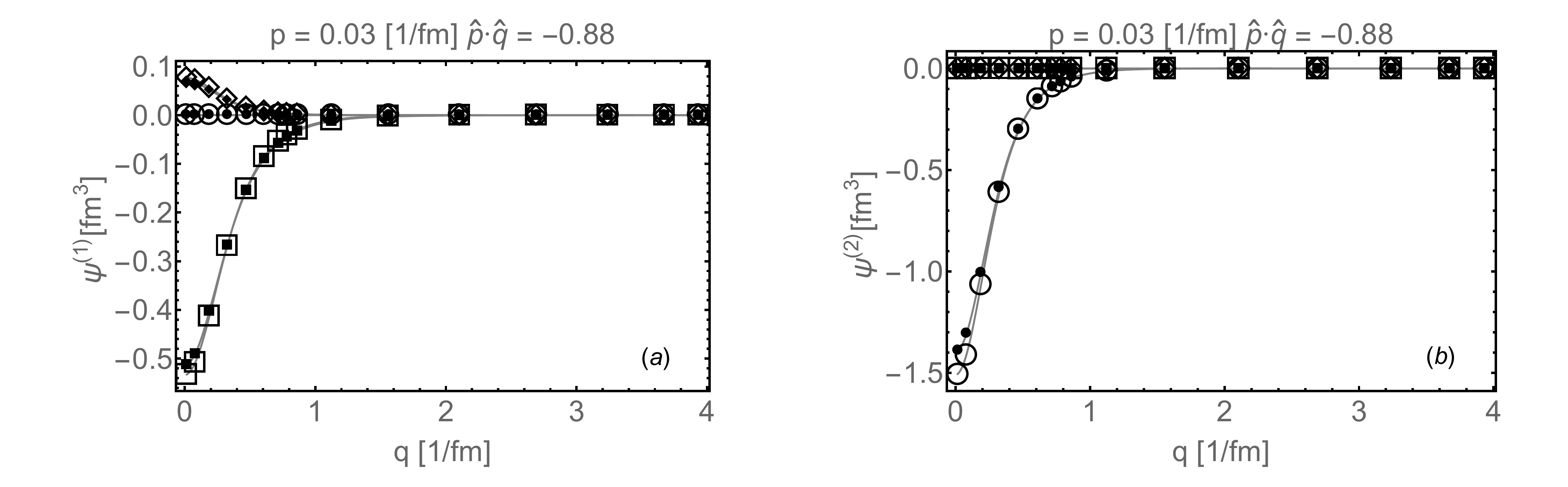}
	\end{center}
	\caption{Similar to Figure \ref{selected_scalar} and \ref{selected_scalar_3H}.
	Comparison of two dominant scalar functions for the
	$^{3}$He Faddeev component calculated with and without the 3N force. Squares circles and diamonds respectively denote the three
	isospin states: $t=0,T=\frac{1}{2}$; 
	$t=1,T=\frac{1}{2}$ and $t=1,T=\frac{3}{2}$. Empty markers correspond to
	$^{3}$He calculated without the 3NF and solid markers correspond to $^{3}$He
	calculated with the 3N force.
	In calculations without the 3N force the eigenvalue from (\ref{eq2}) for energy $E = -7.34124 [\text{MeV}]$ was $\lambda = 0.99997$.
	}
	\label{wave_scalar_3H_c2}
\end{figure}

Tables \ref{avg_3} and \ref{avg_4} contain expectation values for
$^{3}$He and $^{3}$H calculated during the second run and without using a 3N force.
Similarly as in Tab. \ref{avg_1} and \ref{avg_2} the sum of the expectation
values is close to the extrapolated bound state energy and the difference can
be attributed to the small number of grid points used to represent scalar
functions. Additionally, the expectation values of the 3N potential were
calculated for $^{3}$He and $^{3}$H (obtained without using a 3N force). These
values are relatively small and given in the captions of Tab. \ref{avg_3} and
\ref{avg_4}. Finally in Fig. \ref{wave_scalar_3H_c2} two dominant 
scalar functions $\Psi$ for
the $^{3}$He bound state calculated with a 3N force and without a 3N force are
compared. The largest differences appear for the $T = \frac{1}{2}$ components. 

\begin{table}[H]
\[
\begin{array}{cccc}
 <E_{\text{kin}}> & <E_{\text{pot}}^{2N}> & <E_C^{2N}> &
   \Sigma  \\
 30.73 & -38.7 & 0.6953 & -7.268 \\
\end{array}
\]
\caption{Expectation values (in $[\text{MeV}]$) for the 
kinetic energy ($E_{\text{kin}}$), 
the strong 2N potential ($E_{\text{pot}}^{2N}$), 
and the 3N force ($E_{\text{pot}}^{3N}$). These values 
were obtained using the methods outlined in \cite{a_three_dimensional}
from $^{3}$He bound state scalar functions obtained during the 
second run (see text). 
The bound state was calculated using only 2N forces. 
The eigenvalue from (\ref{eq2}) for energy $E = -7.34124 [\text{MeV}]$ was $\lambda = 0.99997$.
Additionally, the expectation value of the 3N force was calculated: $0.03634 [\text{MeV}]$.
}
	\label{avg_3}
\end{table}

\begin{table}[H]
\[
\begin{array}{ccc}
 <E_{\text{kin}}> & <E_{\text{pot}}^{2N}> & \Sigma  \\
 31.32 & -39.29 & -7.973 \\
\end{array}
\]
\caption{Expectation values (in $[\text{MeV}]$) for the 
kinetic energy ($E_{\text{kin}}$), 
the strong 2N potential ($E_{\text{pot}}^{2N}$), 
and the 3N force ($E_{\text{pot}}^{3N}$). These values 
were obtained using the methods outlined in \cite{a_three_dimensional}
from $^{3}$H bound state scalar functions obtained during the 
second run (see text). 
The bound state was calculated using only 2N forces. 
The eigenvalue from (\ref{eq2}) for energy $E = -8.02347 [\text{MeV}]$ was $\lambda = 0.99997$.
Additionally, the expectation value of the screened Coulomb interaction:
$0.7096 [\text{MeV}]$ and 3N force: $0.0578 [\text{MeV}]$ was calculated.
} 
	\label{avg_4}
\end{table}

The obtained $^{3}$H and $^{3}$He wave functions 
were the starting point used to calculate
matrix elements necessary to describe triton beta decay were
calculated for two simple single nucleon current models. Details of this
calculation are available in Appendix \ref{trbetadecay}. Numerical values for the 
matrix elements of the Fermi
current for particle one: 
\begin{equation}
	\BA{\VC{k}'_{1}} \OP{\rho}_{F}(1) \KT{\VC{k}_{1}} = \OP{\tau}(1)_{+} \, \OP{1}
\end{equation}
where $\VC{k'_{1}}$, $\VC{k}$ are the momenta of particle one in the final and
initial states, $\OP{\tau}(1)_{+}$ is the isospin raising operator and $\OP{1}$ is the
identity operator in the spin space of particle one, are gathered in 
Tab. \ref{fcr}. 

Matrix elements for the $z$ component of the Gamow-Teller current:
\begin{equation}
	\BA{\VC{k}'_{1}} \OP{\VC{j}}_{GTz}(1) \KT{\VC{k}_{1}} = \OP{\tau}(1)_{+} \,
	\OP{\sigma}(1)_{z},
\label{jgt}
\end{equation}
where $\OP{\sigma}(1)_{z}$ is a component of the vector spin operator
$\OP{\VC{\sigma}}(1) = (\OP{\sigma}(1)_{x} , \OP{\sigma}(1)_{y} ,
\OP{\sigma}(1)_{z})$,
are gathered in Tab. \ref{gtcr}.
In order to verify these results several additional calculations were performed.
First, matrix elements of (\ref{jgt}) were calculated with 
$\OP{\sigma}(1)_{z}$ replaced
by 
the spherical components of the spin $\OP{\sigma}(1)_{+1}$, $\OP{\sigma}(1)_{-1}$:
\begin{equation}
	\OP{j}_{GT+}(1) = \OP{\tau}(1)_{+} \OP{\sigma}(1)_{+1},
\label{jgtplus}
\end{equation}
\begin{equation}
	\OP{j}_{GT-}(1) = \OP{\tau}(1)_{+} \OP{\sigma}(1)_{-1}.
\label{jgtminus}
\end{equation}
These results are gathered in Tab. \ref{gtcrpm}. 
Since observables are proportional to the sum over the spin projections in the 
initial and final states $m_{i}$, $m_{f}$:
\begin{equation}
S(\OP{j}) = \sum_{m_{f} = \pm \frac{1}{2}} \sum_{m_{i} = \pm \frac{1}{2}} 
|\BA{^{3}\text{He} \,\, m_{f}} \OP{j}(1) \KT{^{3}\text{H}
	\,\, m_{i}}|^{2}
\label{sumj}
\end{equation} 
this sum was calculated for $\OP{\sigma}(1)_{z}$, $\OP{\sigma}(1)_{+1}$ and
$\OP{\sigma}(1)_{-1}$
and compared in Tab. \ref{summm}. The values are essentialy the
same and this confirmes the corectness of the calculations form Tab. \ref{gtcr} and \ref{gtcrpm}.
Next, the expectation values of the $\frac{1}{2}\left(\OP{1} + \OP{\tau}(1)_{3}\right)$
isospin operator ($\OP{\tau}(1)_{3}$ is the third component of the ispspin for
particle $1$) were calculated. The obtained values were $0.666395$ for $^{3}$He and $0.333533$
for $^{3}$H. These two results provide an additional verification since they
are very close to $\frac{2}{3}$ and $\frac{1}{3}$ that
could be expected for $^{3}$He and $^{3}$H respectively. Results for simple single nucleon
currents that are presented here, open up the possibility to perform calculations
with more complicated models and two - nucleon currents. 

\begin{table}[H]
\[
\begin{array}{c|c}
	\text{matrix element:} & \text{value:} \\
	\hline
  \BA{^{3}\text{He} \,\, m_{f} = -\frac{1}{2}} \OP{\rho}_{F}(1) \KT{^{3}\text{H}
	\,\, m_{i} = -\frac{1}{2}} & 0.332825 \\
 \BA{^{3}\text{He} \,\, m_{f} = \frac{1}{2}} \OP{\rho}_{F}(1) \KT{^{3}\text{H} \,\,
	m_{i} = \frac{1}{2}} & 0.332825 \\
\end{array}
\]
\caption{Non zero matrix elements of the Fermi current for particle $1$.
	The spin projection of the triton in the initial state is $m_{i}$
	and the spin projection of helium-$3$ in the final state is $m_{f}$.
} 
	\label{fcr}
\end{table}

\begin{table}[H]
\[
\begin{array}{c|c}
	\text{matrix element:} & \text{value:} \\
	\hline
  \BA{^{3}\text{He} \,\, m_{f} = -\frac{1}{2}} \OP{j}_{GTz}(1) \KT{^{3}\text{H}
	\,\, m_{i} = -\frac{1}{2}} & 0.310749 \\
 \BA{^{3}\text{He} \,\, m_{f} = \frac{1}{2}} \OP{j}_{GTz}(1) \KT{^{3}\text{H} \,\,
	m_{i} = \frac{1}{2}} & -0.310749 \\
\end{array}
\]
\caption{Non zero matrix elements of the Gamow-Teller current for particle $1$.
	The spin projection of the triton in the initial state is $m_{i}$
	and the spin projection of helium-$3$ in the final state is $m_{f}$.
} 
	\label{gtcr}
\end{table}

\begin{table}[H]
\[
\begin{array}{c|c}
	\text{matrix element:} & \text{value:} \\
	\hline
  \BA{^{3}\text{He} \,\, m_{f} = \frac{1}{2}} \OP{j}_{GT+}(1) \KT{^{3}\text{H}
	\,\, m_{i} = -\frac{1}{2}} & 0.439465 \\
 \BA{^{3}\text{He} \,\, m_{f} = -\frac{1}{2}} \OP{j}_{GT-}(1) \KT{^{3}\text{H} \,\,
	m_{i} = \frac{1}{2}} & -0.439465 \\
\end{array}
\]
\caption{Non zero matrix elements of the currents from (\ref{jgtplus}) and (\ref{jgtminus}). 
	The spin projection of the triton in the initial state is $m_{i}$
	and the spin projection of helium-$3$ in the final state is $m_{f}$.
} 
	\label{gtcrpm}
\end{table}

\begin{table}[H]
\[
\begin{array}{c|c}
	\text{sum:} & \text{value:} \\
	\hline
  S(\OP{j}_{GTz}) & 0.19313 \\
  S(\OP{j}_{GT+})& 0.19313 \\
  S(\OP{j}_{GT-})& 0.19313 \\
\end{array}
\]
\caption{
Values of the sum from (\ref{sumj}) for currents (\ref{jgt}), (\ref{jgtplus}) and (\ref{jgtminus}).
The values are essentialy the same for all cases. Since observables are proportional to (\ref{sumj})
(among other things, the proportionality coefficients contain a factor of $3^{2} = 9$ steming from
the necesity to use $\OP{j}(1) + \OP{j}(2) + \OP{j}(3)$) this verifies the
	calculation in Tab. \ref{gtcr} and \ref{gtcrpm}.
} 
	\label{summm}
\end{table}

\section{Summary and outlook}
\label{summary}

This paper shows that three-nucleon bound state calculations with
screened Coulomb potentials are possible using the ``three
dimensional" approach. Instead of relying on the partial wave decomposition of
operators relevant to the calculation, the new approach uses
the three dimensional, momentum, degrees of freedom of the nucleons directly. A practical
numerical realization of these calculations is made
feasible by the idea to write the three-nucleon state as a linear
combination of spin operators and scalar functions. These scalar functions effectively define
the state and are the central object in the ``three dimensional" calculations.

The dependency of the bound state energy on the screening radius of the Coulomb
interaction was shown and it was determined that it is sufficient to perform
calculations with a screening radius of $10 [\text{fm}]$ (for the model of
Coulomb interaction used this means that the potential goes to zero at distances
larger then $30 [\text{fm}]$). At this value the
values of observables are expected to be converged. Plots showing selected scalar functions
were given. Apart from the scalar functions that define the bound state, the
plots also contain additional, overlapping, functions that verify the validity of the
obtained solution. 

Having calculated the bound states of $^{3}$He and $^{3}$H, the expectation
values of the kinetic and potential energies were calculated. This further
confirmed the corectness of the 
obtained results but also showed that it would be beneficial to
perform the calculations using a grater number of grid points to represent
the scalar functions. 
Calculations with an increased number of points are planned for
more modern models of chiral two- and three- nucleon interactions.

Additionally, basic
matrix elements related to the triton beta decay were calculated. These matrix 
elements
are based on simple models of nuclear currents and can be directly used to calculate
observables related to that decay.

\section*{Acknowledgments}
The author would like to thank J. Golak, H. Wita{\l}a, and  
R. Skibi{\'n}ski from the Jagiellonian University for their advice and help in
	preparing this paper as well as A. Nogga from
Forschungszentrum J{\"u}elich for very fruitfull discussions.
The project was financed from the resources of the National
Science Center, Poland, under grants 2016/22/M/ST2/00173 and 2016/21/D/ST2/01120. 
Numerical calculations were preformed on the supercomputer clusters of the
J{\"u}lich Supercomputing Center,
J{\"u}lich, Germany.

\appendix

\section{Spin operators in 3N operator form}
\label{spin_operators}
Below is a list of spin operators used in the operator form of the 3N
(Faddeev,
bound) state (\ref{state_operator_form}). Spin operators acting in the spaces
of particle $1$, $2$, $3$ are denoted as $\OP{\VC{\sigma}}(1)$,
$\OP{\VC{\sigma}}(2)$, $\OP{\VC{\sigma}}(3)$ respectively and $\OP{\VC{\sigma}}(2 , 3) = \frac{1}{2}
(\OP{\VC{\sigma}}(2) - \OP{\VC{\sigma}}(3))$. Vectors $\VC{p}$, $\VC{q}$ are the
Jacobi momenta of the 3N system (if $\VC{k}_{1,2,3}$ are individual particle
momenta then $\VC{p} = \frac{1}{2}(\VC{k}_{2} - \VC{k}_{3})$,
$\VC{q} = \frac{2}{3} (\VC{k}_{1} - \frac{1}{2} (\VC{k}_{2} + \VC{k}_{3}))$).
\begin{align*}
\OP{O}_{1}(\VC{p} , \VC{q}) &= \OP{1} ,\\
\OP{O}_{2}(\VC{p} , \VC{q}) &= \frac{1}{\sqrt{3}} \OP{\VC{\sigma}}(2 , 3) \cdot \OP{\VC{\sigma}}(1),\\
\OP{O}_{3}(\VC{p} , \VC{q}) &= \sqrt{\frac{3}{2}}\frac{1}{i} \OP{\VC{\sigma}}(1) \cdot (\UV{p} \times \UV{q}),\\
\OP{O}_{4}(\VC{p} , \VC{q}) &= \frac{1}{\sqrt{2}} \left( i \OP{\VC{\sigma}}(2 ,
	3) \cdot (\UV{p} \times \UV{q}) - \right. \\ & \left. (\OP{\VC{\sigma}}(1) \times \OP{\VC{\sigma}}(2 , 3)) \cdot (\UV{p} \times \UV{q}) \right),\\
\OP{O}_{5}(\VC{p} , \VC{q}) &= \frac{1}{i} \left(\OP{\VC{\sigma}}(2 , 3) \cdot
	(\UV{p} \times \UV{q}) - \right. \\ & \left.   \frac{i}{2} (\OP{\VC{\sigma}}(1) \times \OP{\VC{\sigma}}(2 , 3)) \cdot (\UV{p} \times \UV{q}) \right),\\
\OP{O}_{6}(\VC{p} , \VC{q}) &= \sqrt{\frac{3}{2}} \left(\OP{\VC{\sigma}}(2 , 3) \cdot \UV{p} \VC{\OP{\sigma}}(1) \cdot \UV{p} - \frac{1}{3} \VC{\OP{\sigma}}(2 , 3) \cdot \VC{\OP{\sigma}}(1) \right),\\
\OP{O}_{7}(\VC{p} , \VC{q}) &= \sqrt{\frac{3}{2}} \left(\OP{\VC{\sigma}}(2 , 3) \cdot \UV{q} \VC{\OP{\sigma}}(1) \cdot \UV{q} - \frac{1}{3} \VC{\OP{\sigma}}(2 , 3) \cdot \VC{\OP{\sigma}}(1) \right),\\
\OP{O}_{8}(\VC{p} , \VC{q}) &= \frac{3}{2} \frac{1}{\sqrt{5}}
	\left(\OP{\VC{\sigma}}(2 , 3) \cdot \UV{q} \VC{\OP{\sigma}}(1) \cdot \UV{p} +
	\right. \\ & \left. \VC{\OP{\sigma}}(2 , 3) \cdot \VC{\OP{p}} \OP{\VC{\sigma}}(1) 
	\cdot \UV{q} - \frac{2}{3} \UV{p} \cdot \UV{q} \VC{\OP{\sigma}}(2 , 3) \cdot \OP{\VC{\sigma}}(1)\right),\\
\end{align*}

\section{Explicit form of operators acting on scalar functions}
\label{explicit_form}

\subsection{Permutation operator}
\label{permutation_operator}

The treatment of the permutation operator in the calculations 
is similar to the
approach that was used in \cite{a_three_dimensional}. The differences
are only formal and related to a slightly different bookkeeping. For this
reason only a sketch of the derivations is provided here. More details
are available in Chapter $10$ of \cite{phd}.

Firstly, the operator form from for the states $\KT{\alpha}$, $\KT{\gamma}$ from
(\ref{state_operator_form}) is inserted into both sides of:
\begin{equation*}
	\left(\OP{1} + \OP{P}\right) \KT{\alpha} = \KT{\gamma}.
\end{equation*}
Next the spin dependency is removed from this equation by projecting it from
the left onto different spin states $\BA{\chi^{m}} \OP{O}_{k}(\VC{p}' \VC{q}')$  for
$k = 1 \ldots 8$ and
summing over $m$.
Finally the resulting integral equations are transformed in
to give the expression for the scalar functions that define $\gamma$. 

Unlike \cite{a_three_dimensional} 
the action of the permutation on the combined isospin - spin state of
the 3N system is considered. Instead of using the $F_{tt'T}$ 
coefficients from \cite{a_three_dimensional} the following functions are
introduced:
\begin{align}
	F^{\text{id}}_{t'T'k;tTi}(\VC{p}',\VC{q'}) = \nonumber \\
	\sum_{m} {\BA{\left(t' \frac{1}{2} \right) T' M_{T'}} \otimes \BA{\chi^{m}}} \MX{\OP{1} \otimes \OP{O}_{k}(\VC{p}' , \VC{q}')} \nonumber \\
\MX{\OP{1} \otimes \OP{O}_{i}(\VC{p}' , \VC{q}')} {\KT{\left(t \frac{1}{2}
	\right) T M_{T}} \otimes \KT{\chi^{m}}},
\end{align}
\begin{align} \nonumber
F_{t'T'k;tTi}^{1223}(\VC{p}',\VC{q'}) = \nonumber \\
	\sum_{m} {\BA{\left(t' \frac{1}{2} \right) T' M_{T'}} \otimes \BA{\chi^{m}}}
	\nonumber \\ \MX{\OP{1} \otimes \OP{O}_{k}(\VC{p}' , \VC{q}')} \MX{\OP{P}_{12}} \MX{\OP{P}_{23}} \nonumber \\
\MX{\OP{1} \otimes \OP{O}_{i}(\VC{P}^{2312}(\VC{p}',\VC{q}') ,
	\VC{Q}^{2312}(\VC{p}',\VC{q}'))} \nonumber \\ {\KT{\left(t \frac{1}{2} \right)
	T M_{T}} \otimes \KT{\chi^{m}}}, 
\end{align}
\begin{align} \nonumber
F_{t'T'k;tTi}^{1323}(\VC{p}',\VC{q'}) = \\ \nonumber
	\sum_{m} {\BA{\left(t' \frac{1}{2} \right) T' M_{T'}} \otimes \BA{\chi^{m}}}
\nonumber \\ 	\MX{\OP{1} \otimes \OP{O}_{k}(\VC{p}' , \VC{q}')} \MX{\OP{P}_{13}} \MX{\OP{P}_{23}} \nonumber \\
\MX{\OP{1} \otimes \OP{O}_{i}(\VC{P}^{2313}(\VC{p}',\VC{q}') ,
	\VC{Q}^{2313}(\VC{p}',\VC{q}'))} \nonumber \\ {\KT{\left(t \frac{1}{2} \right)
	T M_{T}} \otimes \KT{\chi^{m}}},
\end{align}
where $\MX{\OP{P}_{ij}}$ are the matrix representations of the operator
exchanging particles $i$ and $j$ in the combined isospin - spin space of the 3N
system. The functions $\VC{P}^{2313}$, $\VC{P}^{2312}$, $\VC{Q}^{2313}$, and
$\VC{Q}^{2312}$ are a direct result of applying the permutation operator to Jacobi
momentum eignestates:
\begin{align*} 
\OP{P}_{23} \OP{P}_{12} \KT{\VC{p}' \VC{q}'} =
	\KT{\VC{P}^{2312}(\VC{p}',\VC{q}') \VC{Q}^{2312}(\VC{p}',\VC{q}')},\\
\VC{P}^{2312}(\VC{p}',\VC{q}') = - \frac{1}{4} (2 \VC{p}' + 3 \VC{q}'),\\
\VC{Q}^{2312}(\VC{p}',\VC{q}') = \VC{p}' - \frac{1}{2} \VC{q}',\\
\OP{P}_{23} \OP{P}_{13} \KT{\VC{p}' \VC{q}'} =
	\KT{\VC{P}^{2313}(\VC{p}',\VC{q}') \VC{Q}^{2313}(\VC{p}',\VC{q}')}\\
\VC{P}^{2313}(\VC{p}',\VC{q}') = - \frac{1}{4} (2 \VC{p}' - 3 \VC{q}')\\
\VC{Q}^{2313}(\VC{p}',\VC{q}') = - \VC{p}' - \frac{1}{2} \VC{q}'.\\
\end{align*}

If the first part of the permutation operator
$\OP{P} = \OP{P}_{12} \OP{P}_{23} + \OP{P}_{13} \OP{P}_{23}$, namely $\OP{P}_{12}
\OP{P}_{23}$, is applied to a 3N
state $\KT{\alpha}$ written in the form of equation
(\ref{state_operator_form}) ($\KT{\alpha}$ is defined by a set of scalar
functions $\alpha$) then this results in a new 3N state $\KT{\gamma}$ that
can also be written in the form from equation (\ref{state_operator_form}) 
($\KT{\gamma}$ is defined by a set of scalar
functions $\gamma$). This
introduces the operator $\OP{P}^{\mathrm{scalar}}_{1223}$ that is defined via
the relation:
\begin{align} \nonumber
\left(\OP{P}^{\mathrm{scalar}}_{1223} \alpha \right)_{t'T'}^{(k)}(|\VC{p}'| ,
	|\VC{q}'| , \UV{p}' \cdot \UV{q}') = \\ \gamma_{t'T'}^{(k)}(|\VC{p}'| , |\VC{q}'| , \UV{p}' \cdot \UV{q}')
\label{scalargammaphipermutation1223}
\end{align}
or more precisely:
\begin{align} \nonumber
\gamma_{t'T'}^{(k)}(|\VC{p}'| , |\VC{q}'| , \UV{p}' \cdot \UV{q}') =  \nonumber \\
\sum_{tT} \sum_{i = 1}^{8} \phi_{tT}^{(i)}(|\VC{P}^{2312}(\VC{p}',\VC{q}')| ,
	|\VC{Q}^{2312}(\VC{p}',\VC{q}')| , \nonumber \\
	\UV{P}^{2312}(\VC{p}',\VC{q}') \cdot \UV{Q}^{2312}(\VC{p}',\VC{q}')) \nonumber \\
C_{t'T'k;tTi}^{1223}(\VC{p}',\VC{q'}), 
\label{gammapsifinal}
\end{align}
where the functions $C_{t'T'k;tTi}^{1223}$ are:
\begin{align} \nonumber
	C_{t'T'k;tTi}^{1223}(\VC{p}',\VC{q'}) = \\
	\sum_{t''T''j} (F^{\text{id}})^{-1}_{t'T'k;t''T''j}(\VC{p}',\VC{q'})  F^{1223}_{t''T''j;tTi}(\VC{p}',\VC{q'})
\label{C1223}
\end{align}
and 
\begin{align} \nonumber
	\sum_{t''T''j} (F^{\text{id}})^{-1}_{t'T'k;t''T''j}(\VC{p}',\VC{q'})
	F^{\text{id}}_{t''T''j;tTi}(\VC{p}',\VC{q'}) = \\
	\delta_{t' t} \delta_{T' T}
	\delta_{k i}.
	\label{invvv}
\end{align}

The second part of the permutation operator $\OP{P}_{13} \OP{P}_{23}$ can be
introduced in a completely analogous way by replacing everywhere in  
(\ref{scalargammaphipermutation1223}) - (\ref{invvv}) $1223$ by $1323$:
\begin{align} \nonumber
\left(\OP{P}^{\mathrm{scalar}}_{1323} \alpha \right)_{t'T'}^{(k)}(|\VC{p}'| ,
	|\VC{q}'| , \UV{p}' \cdot \UV{q}') = \\
	\gamma_{t'T'}^{(k)}(|\VC{p}'| , |\VC{q}'|
	, \UV{p}' \cdot \UV{q}').
\label{scalargammaphipermutation1323}
\end{align}
Finally, the $\OP{1}$ in $\OP{1} + \OP{P}$ is just the identity operator and
does not change the scalar function. In the end:
\begin{align} \nonumber
	\left(\OP{A}_{\text{1 + P}} \,\, \alpha \right)_{t'T'}^{(k)}(|\VC{p}'| ,
	|\VC{q}'| , \UV{p}' \cdot \UV{q}') = \\
	\nonumber 
	\alpha_{t'T'}^{(k)}(|\VC{p}'| , |\VC{q}'| , \UV{p}' \cdot \UV{q}') + \\ 
	\left(\OP{P}^{\mathrm{scalar}}_{1223} \alpha \right)_{t'T'}^{(k)}(|\VC{p}'| ,
	|\VC{q}'| , \UV{p}' \cdot \UV{q}') + \nonumber \\ \left(\OP{P}^{\mathrm{scalar}}_{1323} \alpha \right)_{t'T'}^{(k)}(|\VC{p}'| ,
	|\VC{q}'| , \UV{p}' \cdot \UV{q}').
\end{align}

A more detailed derivation of these expressions is available in Chapter $10$ of
\cite{phd}. The
numerical implementation of $\OP{P}^{\mathrm{scalar}}_{1223}$ and 
$\OP{P}^{\mathrm{scalar}}_{1323}$ requires the use of interpolations, the
calculations presented in this paper use cubic Hermitian splines.

\subsection{2N potential}
\label{2N_potential}

This section contains expressions necessary to implement
the action of the 2N potential on the scalar functions from the operator form 
(\ref{state_operator_form}). Since the approach used in this section is very
similar to the
methods presented in \cite{a_three_dimensional}, below only a sketch of the
deriviations is presented. 

It is well established (see eg. \cite{the_qm_few_body_problem})
that the matrix element of the 2N force between 
3N Jacobi momentum eigenstates $\BA{\VC{p}'\VC{q}'}$, $\KT{\VC{p}\VC{q}}$ can be written as:
\begin{align} \nonumber
	\BA{\VC{p}' \VC{q}'} \OP{V}
	\KT{\VC{p} \VC{q}}  = \delta^{3}(\VC{q}' - \VC{q}) \nonumber \\
	\sum_{i = 1}^{6} \sum_{t'T'} \sum_{tT} 
	\delta_{tt'} \delta_{M_{T'} M_{T}}
	v^{t'T'tT}_{i}(p' , p , \UV{p}' \cdot \UV{p}) \nonumber \\
	(\OP{1} \otimes \OP{w}_{i}(\VC{p}' , \VC{p}))
	\KT{(t' \frac{1}{2}) T' M_{T'}} \BA{(t \frac{1}{2}) T M_{T}}
	\label{2n_operator_form}
\end{align}
where $\OP{1} \otimes \OP{w}_{i}(p' , p , \UV{p}' \cdot \UV{p})$ is a spin operator
with $\OP{w}_{i}(p' , p , \UV{p}' \cdot \UV{p})$ acting in the space of particles $2$ and $3$.
Any operator that can be written using (\ref{2n_operator_form}) implicitly satisfies symmetries with respect
to spatial rotations, parity inversion, time reversal and particle exchange and
is effectively defined by the set of scalar functions of the relative final and
initial momenta $v^{t'T'tT}_{i}(p' , p , \UV{p}' \cdot \UV{p})$. The spin
operators $\OP{w}_{i}(\VC{p}' , \VC{p})$ from (\ref{2n_operator_form}) are
\cite{two_nucleon_systems,the_qm_few_body_problem}:
\begin{align*}
	\OP{w}_{1}(\VC{p}' , \VC{p}) = \OP{1} \\
	\OP{w}_{2}(\VC{p}' , \VC{p}) = \OP{\VC{\sigma}}(1) \cdot \OP{\VC{\sigma}}(2) \\
	\OP{w}_{3}(\VC{p}' , \VC{p}) = -i (\OP{\VC{\sigma}}(1) + \OP{\VC{\sigma}}(2)) \cdot (\VC{p}' \times \VC{p})\\
	\OP{w}_{4}(\VC{p}' , \VC{p}) = \OP{\VC{\sigma}}(2) \cdot (\VC{p}' \times \VC{p}) \OP{\VC{\sigma}}(3) \cdot (\VC{p}' \times \VC{p})\\
	\OP{w}_{5}(\VC{p}' , \VC{p}) = \OP{\VC{\sigma}}(2) \cdot (\VC{p}' + \VC{p}) \OP{\VC{\sigma}}(3) \cdot (\VC{p}' + \VC{p})\\
	\OP{w}_{6}(\VC{p}' , \VC{p}) = \OP{\VC{\sigma}}(2) \cdot (\VC{p}' \times \VC{p}) \OP{\VC{\sigma}}(3) \cdot (\VC{p}' - \VC{p})\\
\end{align*}
and are constructed from the relative momentum vector operators $\VC{p}'$, $\VC{p}$ and the spin vector 
operators $\OP{\VC{\sigma}}(i)$ acting
in the space of particles $i = 1 , 2$. 

In the previous paper
\cite{a_three_dimensional} the operator form (\ref{2n_operator_form}) was used
to work out the numerical 
implementation of $\OP{G}_{0}(E) \OP{V}$ in ``three dimensional" bound state calculations.
This was achieved by plugging the operator form (\ref{state_operator_form}) of 3N states $\KT{\alpha}$,
$\KT{\gamma}$, the 2N
potential (\ref{2n_operator_form}) and the momentum space expression
for the free propagator:
\begin{align} \nonumber
	\BA{\VC{p}' \VC{q}'} \OP{G}_{0}(E) \KT{\VC{p} \VC{q}} = \delta^{3}(\VC{p} -
	\VC{p}') \delta^{3}(\VC{q} - \VC{q}') \nonumber \\
	\frac{1}{E - \frac{3}{4 m} p^{2} -
	\frac{1}{2 m} q^{2}}
	\label{free_propagator_momentum_space}
\end{align}
into
\begin{equation*}
	\OP{G}_{0}(E) \OP{V} \KT{\alpha} = \KT{\gamma},
\end{equation*}
and projecting the
resulting equation onto a Jacobi momentum eigenstate $\BA{\VC{p}' \VC{q}'}$.
The spin dependency of the resulting equation was removed by projecting it from
the left onto different spin states $\BA{\chi^{m}} \OP{O}_{k}(\VC{p}' \VC{q}')$  for
$k = 1 \ldots 8$ and
summing over $m$. 
The result of these manipulations is a linear relation:
\begin{equation}
\OP{A}_{G_{0} V}(E) \alpha \equiv \gamma.
\label{a_e_2n}
\end{equation}
that defines the energy $E$ dependant operator $\OP{A}_{G_{0} V}(E)$. It acts in the space
of the set of scalar functions  
that define Faddeev component in
(\ref{state_operator_form}). 
Applying the first part of the Faddeev equation (\ref{faddeq}),
$\OP{G}_{0}(E) \OP{V}$, to a state $\KT{\alpha}$ written in the operator form (\ref{state_operator_form}) 
that is defined by a set of scalar functions $\alpha$ ($\alpha^{(k)}_{t'T'}(p' , q' , x' = \UV{p}'
	\cdot \UV{q}')$) results in a new state $\KT{\gamma}$ that can be written
in the same form (\ref{state_operator_form}) but with a different set of scalar
functions $\gamma$ ($\gamma^{(k)}_{t'T'}(p' , q' , x' = \UV{p}'
	\cdot \UV{q}')$). The full form of $\OP{A}_{G_{0} V}(E)$ is:
\begin{align} \nonumber
\left( \OP{A}_{G_{0} V}(E) \alpha \right)^{(k)}_{t'T'}(p' , q' , x' = \UV{p}'
	\cdot \UV{q}') \equiv \\ \nonumber
\gamma^{(k)}_{t'T'}(p' , q' , x') = \\ \nonumber 
\int_{0}^{\infty} \text{d}p p^{2} \int_{-1}^{1} \text{d}x \sum_{T} \sum_{i = 1}^{8}
\frac{1}{E - \frac{3}{4 m} p'^{2} - \frac{1}{m} q'^{2}} \\
\left\{\sum_{j = 1}^{6} \int_{0}^{2 \pi} \text{d}\phi \, v_{j}^{t'T't'T}(p' , p
	, \UV{p}' \cdot \UV{p}) \bar{L}_{kji}(\VC{p}' , \VC{q}' , \VC{p}) \right\}
	\nonumber \\
	\alpha^{(i)}_{t'T}(p , q' , x),
\label{2n_scalar_operator}
\end{align}
where the integral over the vector $\VC{p}$ is parametrized as:
\begin{equation*}
	\VC{p} = p (\sqrt{1 - x^{2}} \cos{\phi} , \sqrt{1 - x^{2}} \sin{\phi} , x)
\end{equation*}
with $p$ being the magnitude of $\VC{p}$, $\phi$ being the azimutal
angle and $x$ being the cosine of the polar angle.  
Additionally, using the scalar character of the equations, 
in (\ref{2n_scalar_operator}):
\begin{align*}
\VC{p}' = p' (\sqrt{1 - x'^{2}} , 0 , x'), \\
\VC{q}' = q' (0 , 0 , 1) \\
\end{align*}
can be chosen.

The curly brackets in (\ref{2n_scalar_operator}) contain integrals 
\begin{align} \nonumber
I^{\text{2N}}_{t'T'T;k;i}(p' , q' , x' , p , x) = \\ 
	\sum_{j = 1}^{6} \int_{0}^{2 \pi} \text{d}\phi \, v_{j}^{t'T't'T}(p' , p , \UV{p}' \cdot \UV{p}) \bar{L}_{kji}(\VC{p}' , \VC{q}' , \VC{p})
\label{2N_integral}
\end{align}
that can be performed once and reused later
when applying $\OP{G}_{0} \OP{V}$ to different states 
(which are defined by different scalar functions $\alpha$). The $\bar{L}$
functions are defined as:
\begin{equation*}
\bar{L}_{kji} \equiv \sum_{l} C^{-1}_{kl} L_{lji}
\end{equation*}
where the $C^{-1}$, $L$ coefficients are defined in equations (49), (30) from \cite{a_three_dimensional}.

\begin{figure}[H]
	\begin{center}
		\includegraphics[width = 0.8 \textwidth]{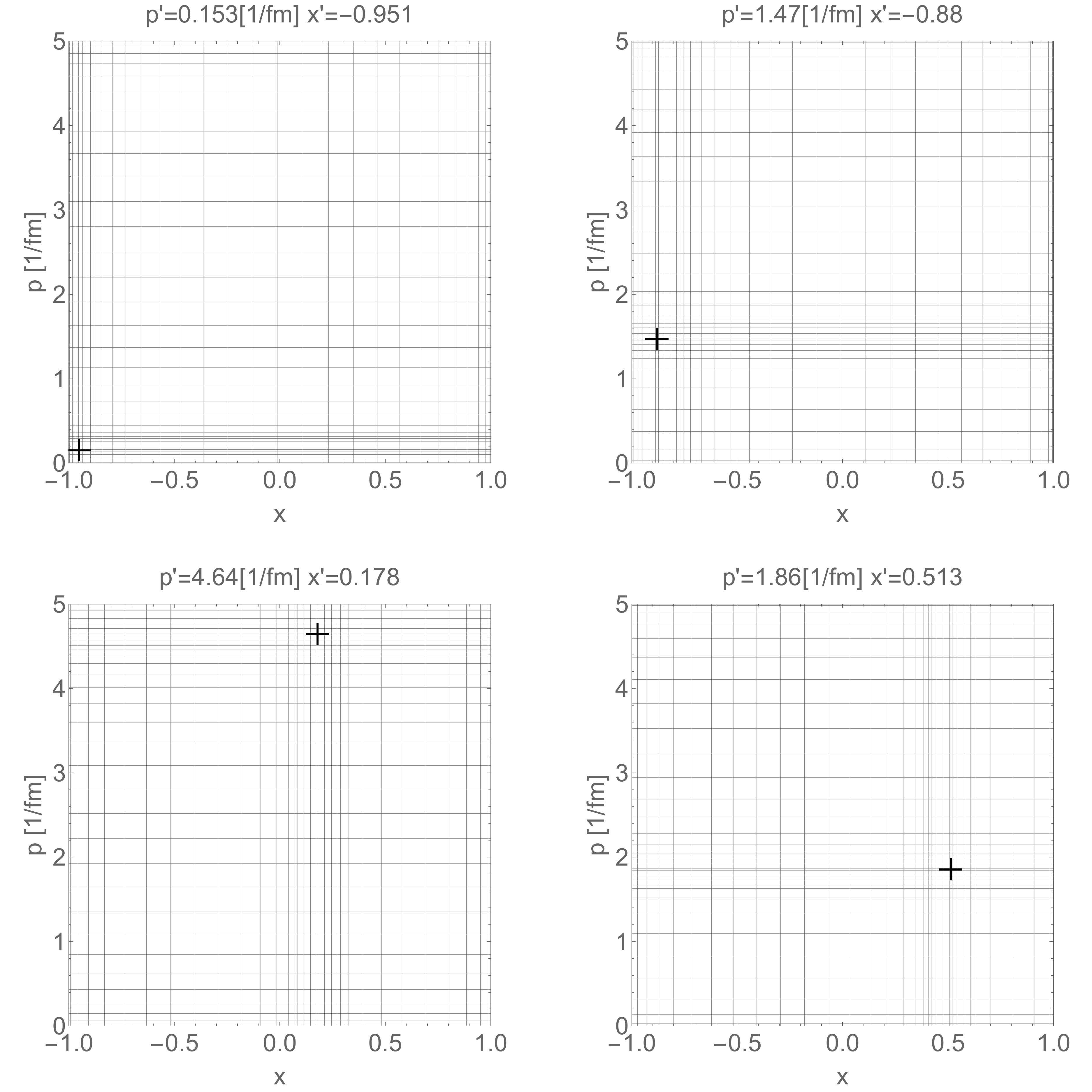}
	\end{center}
	\caption{Integration points used for the screened Coulomb interaction in
	(\ref{2n_scalar_operator}). Grid lines correspond to integration points in the $x$
	and $p$ directions. The cross corresponds to $(x' , p')$. Integration points
	in each direction are created by stiching togather several sets of Gaussian
	quadrature points.}
	\label{coulomb_integration_points}
\end{figure}

Equation (\ref{2n_scalar_operator}) can be used to perform numerical calculations 
with both the short range and long range  2N interactions. When calculating the bound state 
of $^{3}$He with the screened Coulomb interaction, the operator 
$\OP{A}_{G_{0} V}$ is further split into two parts:
\begin{equation*}
	\OP{A}_{G_{0} V} = \OP{A}_{G_{0 }V_{NN}} + \OP{A}_{G_{0} V_{C}}
\end{equation*}
that correspond to the short range nuclear interaction and longer ranged
screened Coulomb force.
The numerical implementation of both of these operators is very similar. The only practical difference
is that in the long range part $\OP{A}_{G_{0} V_{C}}$ the integrals 
$I^{\text{2N}}_{t'T'T;k;i}(p' , q' , x' , p , x)$ are calculated using more integration points 
around $p' = p$ and $x' = x$. Examples of grid points in the $p - x$ plane
for different values of $p'$, $x'$ are shown in Fig.
\ref{coulomb_integration_points}. Also the number of integration points along
the azimuthal angle in (\ref{2N_integral}) is increased, typically $256$
Gaussian points were used compared to $16$ and $64$ points for the nuclear
interaction in the first and second run of the calculations.

For the results presented in this paper a screened Coulomb potential from 
Ref. \cite{screened_coulomb} was used. The momentum space expressions for this 
interaction are given
explicitly in equation (A3) and (A4) in Ref. \cite{screened_coulomb}. 
Solutions were obtained using the same, first generation, short range NNLO 2N interaction 
as was used in \cite{a_three_dimensional} but use both the neutron-proton and
proton-proton version of this interaction (the neutron-neutron interaction is
approximated by the proton-proton force).

\subsection{3N force}
\label{3N_force}

This section contains expressions necesary to implement the
action of the 3N potential on the scalar functions from the operator form
(\ref{state_operator_form}).
The matrix element of the 3N force between 3N Jacobi momentum
eigenstates $\BA{\VC{p}'\VC{q}'}$, $\KT{\VC{p}\VC{q}}$ can be written 
\cite{a_three_dimensional} as:
\begin{align} \nonumber
	\BA{\VC{p}' \VC{q}'} \OP{V}^{(1)} \KT{\VC{p} \VC{q}}  = \\ \nonumber 
	\sum_{t'T'} \sum_{tT} 
	\delta_{T T'} \delta_{M_{T} M_{T'}} 
	\OP{V}^{(1)}_{tt'T}(\VC{p}',\VC{q}',\VC{p}, \VC{q}) \\
	\KT{(t' \frac{1}{2}) T' M_{T'}} \BA{(t \frac{1}{2}) T M_{T}}
	\label{3n_operator_form}
\end{align}
where $\OP{V}^{(1)}_{tt'T}(\VC{p}',\VC{q}',\VC{p}, \VC{q})$ is a spin operator acting in the space of 
three particles. The momentum dependece of this spin operator is limited by the
requirement of spatial rotation invariance. Using this limitation the
$\OP{V}^{(1)}_{tt'T}(\VC{p}',\VC{q}',\VC{p}, \VC{q})$
can be written as a linear combination of $64$ scalar functions and known spin
operators (for more details see Appendix A in \cite{general_operator_form}).
Unfortunately using this operator form does not directly impact numerical
performance and for this reason the general operator form was not used and 
a new method of carrying out the necessary
numerical integrations was developed instead.
However, having a uniform template for the 3NF can be very beneficial when testing
new models of nuclear interactions and is strongly encouraged. 
Inserting the general operator form of the 3NF from \cite{general_operator_form}
into three dimensional bound state calculations is an easy task.
The results presented in this paper use the same first generation, NNLO 3N force 
as was used in \cite{a_three_dimensional}. 

In \cite{a_three_dimensional} the implementation of the second
term of the Faddeev equation (\ref{faddeq}), $\OP{G}_{0}(E)
\OP{V}^{(1)}$, in ``three dimensional"
calculations was worked out. Similarly as in section \ref{2N_potential} the
implementation is an energy $E$ dependent linear operator $\OP{A}^{\text{3N}}(E)$ 
acting in the space of scalar functions that
are used to define states in (\ref{state_operator_form}).
This operator $\OP{A}_{G_{0} V^{(1)}}(E)$ is defined via the
relation:
\begin{equation}
	\OP{A}_{G_{0} V^{(1)}}(E) \alpha \equiv \gamma,
\end{equation}
where $\alpha$ ($\alpha^{(k)}_{t'T'}(p' , q' , x' = \UV{p}' \cdot \UV{q}')$) are
scalar functions that together with (\ref{state_operator_form}) can be used to
reproduce the 3N state $\KT{\alpha}$ before the application of $\OP{G}_{0}(E)
\OP{V}^{(1)}$ and $\gamma$ ($\gamma^{(k)}_{t'T'}(p' , q' , x' = \UV{p}'
\cdot \UV{q}')$) are scalar functions that together with
(\ref{state_operator_form}) can be used to reproduce the 3N state $\KT{\gamma}$ after
the application of $\OP{G}_{0}(E) \OP{V}^{(1)}$. 
The full form of this relation is given explicitly in
\cite{a_three_dimensional} and here we give only a simplified version:
\begin{align} \nonumber
	\left( \OP{A}^{\text{3N}}(E) \alpha \right)^{(l)}_{t'T'}(p' , q' , \UV{p}'
	\cdot \UV{q}') \equiv \\ \nonumber 
	\gamma^{(l)}_{t'T'}(p' , q' , \UV{p}'
	\cdot \UV{q}') = \\ \nonumber
	\int \text{d}^{3}\VC{p} \text{d}^{3}\VC{q} \frac{1}{E - \frac{3}{4 m} p'^{2} -
	\frac{1}{m} q'^{2}} \\ \sum_{t} \sum_{i = 1}^{8}
	\bar{E}_{li}^{t'tT'}(\VC{p'} , \VC{q'} , \VC{p} , \VC{q}) \alpha^{(i)}_{tT'}(p , q , \UV{p} \cdot \UV{q})
	\label{3NF_implementation}
\end{align}
where the $\bar{E}_{li}^{t'tT'}$ functions are expressed in terms of the
$C^{-1}$ and $E$ coefficients from Eqs. (49) and (27) in
\cite{a_three_dimensional}:
\begin{equation*}
\bar{E}_{ki} = \sum_{l} C^{-1}_{kl} E_{li}.
\end{equation*}
Equation (\ref{3NF_implementation}) is obtained by plugging the operator form of
the Faddeev component (\ref{state_operator_form}), the 3N force
(\ref{3n_operator_form}) and the free propagator
(\ref{free_propagator_momentum_space}) into 
\begin{equation*}
	\OP{G}_{0}(E) \OP{V}^{(1)} \KT{\alpha} = \KT{\gamma}
\end{equation*}
and removing the spin dependencies 
by projecting it from
the left onto a Jacobi momentum eigenstate $\BA{\VC{p}' \VC{q}'}$, different spin states $\BA{\chi^{m}} \OP{O}_{k}(\VC{p}' \VC{q}')$  for
$k = 1 \ldots 8$ and
summing over $m$.

It turns out that the parametrization of the six fold integral in equation
(\ref{3NF_implementation}) is crucial to the numerical efficiency of the
calculation. In \cite{a_three_dimensional} the following parametrization of the
$\VC{p}$ and $\VC{q}$ vectors was used:
\begin{align} \nonumber
	\VC{p} = p \left(\cos{\phi_{\VC{p}}} \sqrt{1 - x_{\VC{p}}^{2}} , \sin{\phi_{\VC{p}}} \sqrt{1 -
	x_{\VC{p}}^{2}} , x_{\VC{p}} \right) = \\ p \OP{R}_{\UV{z}}^{\phi_{\VC{p}}} \OP{R}_{\UV{y}}^{\theta_{\VC{p}}} \left(0 , 0 , 1\right)
	\label{p_parametrization}
\end{align}
\begin{align} \nonumber
	\VC{q} = q \left(\cos{\phi_{\VC{q}}} \sqrt{1 - x_{\VC{q}}^{2}} , \sin{\phi_{\VC{q}}} \sqrt{1 -
	x_{\VC{q}}^{2}} , x_{\VC{q}} \right) = \\ q \OP{R}_{\UV{z}}^{\phi_{\VC{q}}}
	\OP{R}_{\UV{y}}^{\theta_{\VC{q}}} \left(0 , 0 , 1\right),
	\label{old_q_parametrization}
\end{align}
where $\OP{R}_{\UV{e}}^{\alpha}$ is a spatial rotation around the unit vector
$\UV{e}$ by angle $\alpha$, $p$ ($q$) is the magnitude of the momentum
vector $\VC{p}$ ($\VC{q}$), $\phi_{\VC{p}}$ ($\phi_{\VC{q}}$) is the azimuthal
angle of vector $\VC{p}$ ($\VC{q}$) and $x_{\VC{p}}$ ($\VC{x}_{\VC{q}}$) is the
cosine of the polar angle of vector $\VC{p}$ ($\VC{q}$). This parametrization
results in the following form of
equation (\ref{3NF_implementation}):
\begin{align} \nonumber
	\gamma^{(l)}_{t'T'}(p' , q' , \UV{p}' \cdot \UV{q}') = \\ \nonumber 
	\int_{0}^{\infty} \text{d}p \, p^{2} \int_{-1}^{1} \text{d}x_{\VC{p}} \int_{0}^{2 \pi} \text{d}\phi_{\VC{p}} 
	\int_{0}^{\infty} \text{d}q \, q^{2} \\ \nonumber \int_{-1}^{1} \text{d}x_{\VC{q}} \int_{0}^{2 \pi}
	\text{d}\phi_{\VC{q}} \frac{1}{E - \frac{3}{4 m} p'^{2} - \frac{1}{m} q'^{2}}
	\\ \nonumber \sum_{t} \sum_{i = 1}^{8}
	\bar{E}_{li}^{t'tT'}(\VC{p'} , \VC{q'} , \VC{p} , \VC{q}) \alpha^{(i)}_{tT'}(p , q , \UV{p} \cdot \UV{q})
\end{align}
where the angle argument of the scalar function $\alpha^{(i)}_{tT'}(p , q , \UV{p}
\cdot \UV{q})$ has a complicated form:
\begin{align} \nonumber
	\UV{p} \cdot \UV{q} = \\
	\cos{\phi_{\VC{p}}} \cos{\phi_{\VC{q}}} + \cos{(\phi_{\VC{p}} -
	\phi_{\VC{q}})} \sqrt{1 - x_{\VC{p}}^{2}} \sqrt{1 - x_{\VC{q}}^{2}})
\end{align}
and the many-fold integral has to be calculated each time $\OP{G}_{0}(E)
\OP{V}^{(1)}$ is applied.
In the present work we use a different parametrization of the vector $\VC{q}$:
\begin{align} \nonumber
	\VC{q} = q \OP{R}_{\UV{z}}^{\phi_{\VC{p}}} \OP{R}_{\UV{y}}^{\theta_{\VC{p}}}
	\OP{R}_{\UV{z}}^{\phi_{\VC{q}}} \OP{R}_{\UV{y}}^{\theta_{\VC{q}}} \left(0 , 0
	, 1\right) = \\
	\nonumber
	(\sqrt{1-x_{\VC{p}}^2} x_{\VC{q}} \cos (\phi_{\VC{p}})+ \\ \nonumber \sqrt{1-x_{\VC{q}}^2} (x_{\VC{p}}
   \cos (\phi_{\VC{p}}) \cos (\phi_{\VC{q}})- \\ \nonumber \sin (\phi_{\VC{p}}) \sin
	 (\text{$\phi_{\VC{q}}
   $})), \\ \nonumber
	 \sqrt{1-x_{\VC{p}}^2} x_{\VC{q}} \sin (\phi_{\VC{p}})+ \\ \nonumber \sqrt{1-x_{\VC{q}}^2}
   (x_{\VC{p}} \sin (\phi_{\VC{p}}) \cos (\phi_{\VC{q}})+ \\ \nonumber \cos (\phi_{\VC{p}}) \sin
   (\phi_{\VC{q}})), \\ 
	 x_{\VC{p}} x_{\VC{q}}-\sqrt{1-x_{\VC{p}}^2} \sqrt{1-x_{\VC{q}}^2} \cos(\phi_{\VC{q}})).
	 \label{q_parametrization}
\end{align}
The new parametrization (\ref{q_parametrization}) is related to the old
parametrization (\ref{old_q_parametrization}) by a spatial rotation and the
absolute value of the Jacobian determinant of the coordinate transformation from (\ref{p_parametrization}),
(\ref{old_q_parametrization}) to (\ref{p_parametrization}),
(\ref{q_parametrization}) is $1$. Using
(\ref{q_parametrization}) together with (\ref{p_parametrization}) leads to a very
simple form of the third argument of the scalar functions:
\begin{equation}
	\UV{p} \cdot \UV{q} = x_{\VC{q}}
\end{equation}
and allows Eq. (\ref{3NF_implementation}) to be written as:
\begin{align} \nonumber
	\gamma^{(l)}_{t'T'}(p' , q' , \UV{p}' \cdot \UV{q}' = x') = \\ \nonumber 
	\int_{0}^{\infty} \text{d}p \, p^{2} \int_{0}^{\infty} \text{d}q \, q^{2}
	\int_{-1}^{1} \text{d}x_{\VC{q}} \\ \nonumber \frac{1}{E - \frac{3}{4 m} p'^{2} - \frac{1}{m} q'^{2}}
	\sum_{t} \sum_{i = 1}^{8} \\ \nonumber
	\left\{ \int_{0}^{2 \pi} \text{d}\phi_{\VC{p}} \int_{-1}^{1} \text{d}x_{\VC{p}} 
	\int_{0}^{2 \pi}
	\text{d}\phi_{\VC{q}} 
	\bar{E}_{li}^{t'tT'}(\VC{p'} , \VC{q'} , \VC{p} , \VC{q}) \right\} \\ 
	\alpha^{(i)}_{tT'}(p , q , \UV{p} \cdot \UV{q} = x_{\VC{q}})
\end{align}
where the integrals in the curly brackets:
\begin{align} \nonumber
	I^{\text{3N}}_{t;i}(p' , q' , x', p , q , x_{\VC{q}}) = \\ \nonumber \int_{0}^{2 \pi} \text{d}\phi_{\VC{p}}
	\int_{-1}^{1} \text{d}x_{\VC{p}} 
	\int_{0}^{2 \pi}
	\text{d}\phi_{\VC{q}} 
	\bar{E}_{li}^{t'tT'}(\VC{p'} , \VC{q'} , \VC{p} , \VC{q})
	\label{3N_integral}
\end{align}
can be performed once, stored in arrays and reused. The possibility to
reuse these integrals significantly reduces the numerical work needed to carry
out the bound state calculations, especially if the calculations are to be performed for a
vast spectrum of bound state energy candidates $E$. The downside of this
approach is that that $I^{\text{3N}}_{t;i}(p' , q' , x', p , q , x_{\VC{q}})$
must be calculated and stored for a large number of parameters - the floating
point parameters $p' , q' , x', p , q , x_{\VC{q}}$ and discrete indeces $t, i$.

The same approach to changing the integration variable can also be applied to
calculations of the expectation value of the 3N force. The
integrations neccesary to carry out this calculation are outlined in
\cite{a_three_dimensional} and are similar to those used in $\OP{A}_{G_{0}
V^{(1)}}$. This allowed the expectation value to be quickly calculated for a
number of different 3N states.

\section{Matrix elements related to the triton beta decay}
\label{trbetadecay}

The calculation of observables related to the beta decay process:
\[
	^{3}\text{H} \rightarrow ^{3}\text{He} + e + \bar{\nu}_{e}
\]
requires computing matrix elements of the form:
\begin{equation}
	\BA{\Psi_{^{3}\text{He}}} \OP{j}(i) \KT{\Psi_{^{3}\text{H}}}
	\label{mej}
\end{equation}
where $\OP{j}(i = 1 , 2 , 3)$ is the single nucleon current acting in the space
of nucleon $i$. In this paper two simple current models are used:
\begin{equation}
	\BA{\VC{k}_{i}'} \OP{j}_{GT}(i) \KT{k_{i}} = \OP{\tau}_{i}^{+}
	\OP{\sigma}(i)_{z} :=\OP{\tau}_{i}^{+} \OP{f}_{GT}(\VC{k}_{i}' , \VC{k_{i}}),
	\label{jGT}
\end{equation}
and
\begin{equation}
	\BA{\VC{k}_{i}'} \OP{\rho}_{F}(i) \KT{\VC{k}_{i}} = \OP{\tau}_{i}^{+}
	\OP{1} := \OP{\tau}_{i}^{+} \OP{f}_{F}(\VC{k}_{i}' , \VC{k_{i}}),
	\label{jF}
\end{equation}
where on the right hand side the dependence on single partical momentum in the
initial and final state was added for generality.
In (\ref{jGT}) and (\ref{jF}) $\OP{\tau}_{i}^{+}$ is the isospin raising operator for
particle $i$, $\OP{\sigma}(i)_{z}$ is the $z$ component of the spin operator for
particle $i$, $\OP{1}$ is the identity operator in spin space and $\VC{k}_{i}'$, $\VC{k}_{i}$
are the momenta of particle $i$ in the final and initial states.  

Using the property:
\begin{align*}
	\BA{\VC{k}_{1}', \VC{k}_{2}', \VC{k}_{3}'} \OP{j}(1) \KT{\VC{k}_{1},
	\VC{k}_{2}, \VC{k}_{3}} = \\
	\OP{f}(\VC{k}_{1}' , \VC{k}_{1}) \delta^{3}(\VC{k}_{2}' - \VC{k}_{2})
	\delta^{3}(\VC{k}_{3}' - \VC{k}_{3}),
\end{align*}
and the relation between the single particle momentum eigenstates and
Jacobi momentum eigenstates: 
\begin{align*}
	\BK{\VC{k}_{1} \VC{k}_{2} \VC{k}_{3}}{\VC{p} \VC{q} \VC{K}} = 
	\delta^{3}\left(\VC{p} - \frac{1}{2} (\VC{k}_{2} - \VC{k}_{3})\right) \\
	\delta^{3}\left(\VC{q} - \frac{2}{3} \left(\VC{k}_{1} - \frac{1}{2}
	(\VC{k}_{2} + \VC{k}_{3})\right) \right) \\
	\delta^{3}\left(\VC{K} - \VC{k}_{1} - \VC{k}_{2} - \VC{k}_{3} \right)
\end{align*}
where $\VC{K}$ is the total momentum of the 3N system,
the operator form of the $^{3}$He and triton bound states from
(\ref{state_operator_form}), the assumption that the total momentum of the
3N system does not change in the triton beta decay process
and that initialy the triton is at rest
it is possible to write (\ref{mej}) for particle $i = 1$ as:
\begin{align} \nonumber
	\BA{\Psi_{^{3}\text{He}} , m_{f}} \OP{j}(1) \KT{\Psi_{^{3}\text{H}} , m_{i}} = \\ \nonumber
	\int \text{d}^{3}\VC{p} \text{d}^{3}\VC{q} \sum_{t' T'} \sum_{t T} 
	\BA{(t' \frac{1}{2}) T' M_{T'}} \OP{\tau}_{1}^{+} \KT{(t \frac{1}{2}) T M_{T}}
	\\ \nonumber
	\sum_{i = 1}^{8} \sum_{j = 1}^{8} 
	\Psi^{(i)}_{^{3}\text{He}; t'T'}(p , q , \UV{p} \cdot
	\UV{q})\Psi^{(j)}_{^{3}\text{H}; tT}(p , q , \UV{p} \cdot \UV{q}) \\ \nonumber
	\BA{\chi_{m_{f}}} 
	\OP{O}_{j}(\VC{p}' , \VC{q}')^{\dagger} 
	\OP{f}(\VC{k}_{1}' = \VC{q},
	\VC{k}_{1} = \VC{q}) \\
	\OP{O}_{i}(\VC{p}' , \VC{q}') \KT{\chi_{m_{i}}}
	\label{current_integral}
\end{align}
where $m_{f}$, $m_{i}$ are spin projections of the 3N system in the final and
initial state and $\Psi_{^{3}\text{He}}$, $\Psi_{^{3}\text{H}}$ are scalar
functions that determine the $^{3}$He and $^{3}$H bound states respectively.

The integral in (\ref{current_integral}) can be easily calculated using a
similar choice of integral variables as in \ref{3N_force}:
\begin{equation*}
	\VC{p} = p \OP{R}_{\UV{z}}^{\phi_{\VC{p}}} \OP{R}_{\UV{y}}^{\theta_{\VC{p}}}
	\left(0 , 0
	, 1\right),
\end{equation*}
\begin{equation*}
	\VC{q} = q \OP{R}_{\UV{z}}^{\phi_{\VC{p}}} \OP{R}_{\UV{y}}^{\theta_{\VC{p}}}
	\OP{R}_{\UV{z}}^{\phi_{\VC{q}}} \OP{R}_{\UV{y}}^{\theta_{\VC{q}}} \left(0 , 0
	, 1\right).
\end{equation*}
Chosing this parametrization leads to $\UV{p} \cdot \UV{q} = \cos(\theta_{q}) :=
x_{\VC{q}}$.
This can be used to greatly simplify (\ref{current_integral}) to the
following expression:
\begin{align} \nonumber
	\BA{\Psi_{^{3}\text{He}} , m_{f}} \OP{j}(1) \KT{\Psi_{^{3}\text{H}} , m_{i}} = \\ \nonumber
	\int_{0}^{\infty} \text{d}p p^{2} \int_{0}^{\infty} \text{d}q q^{2}
	\int_{-1}^{1} \text{d}x_{\VC{q}} \\ \nonumber
	\sum_{t' T'} \sum_{t T} \BA{(t' \frac{1}{2}) T' M_{T'}} \OP{\tau}_{1}^{+}
	\KT{(t \frac{1}{2}) T M_{T}} \\ \nonumber 
	\sum_{i = 1}^{8} \sum_{j = 1}^{8} \Psi^{(i)}_{^{3}\text{He};t'T'}(p , q , x_{\VC{q}})
	\Psi^{(j)}_{^{3}\text{H};tT}(p , q , x_{\VC{q}}) \\ \nonumber
	\left\{ \int_{0}^{\pi} \text{d}\theta_{\VC{p}} \sin(\theta_{\VC{p}}) \int_{0}^{2 \pi} \text{d}\phi_{\VC{p}} \int_{0}^{2 \pi} \text{d}\phi_{\VC{q}}
	\BA{\chi_{m_{f}}} 
	\OP{O}_{j}(\VC{p} , \VC{q})^{\dagger} \right. 
	\\ \nonumber
	\left. 
	\vphantom{\int_{0}^{\pi}}
	\OP{f}(\VC{k}_{1}' = \VC{q},
	\VC{k}_{1} = \VC{q}) 
	\OP{O}_{i}(\VC{p} , \VC{q}) \KT{\chi_{m_{i}}} \right\}
\end{align}
where the integrals in the curly brackets can be computed once and then
used in calculations with different scalar functions $\Psi$.

\end{document}